\documentclass{aastex63}

\newcommand\aastex{AAS\TeX}

\received{April 30, 2020}
\revised{July 21, 2020}
\accepted{August 10, 2020}

\submitjournal{ApJ}

\usepackage{comment}
\usepackage{amsmath}
\usepackage{ulem}

\shorttitle{\aastex\ draft}
\shortauthors{Ikuta et al.}

\renewcommand{\baselinestretch}{1.3}

\begin{document}

\title{Starspot mapping with adaptive parallel tempering I: Implementation of computational code }

\correspondingauthor{Kai Ikuta}
\email{ikuta@kusastro.kyoto-u.ac.jp}

\author[0000-0002-5978-057X]{Kai Ikuta}
\affil{Department of Astronomy, Kyoto University, Kitashirakawa-Oiwake-cho, Sakyo, Kyoto, Kyoto 606-8502, Japan}

\author[0000-0003-0332-0811]{Hiroyuki Maehara}
\affil{Okayama Branch Office, Subaru Telescope, National Astronomical Observatory of Japan, NINS, Kamogata, Asakuchi, Okayama 719-0232, Japan}
\affil{Astronomical Observatory, Kyoto University, Kitashirakawa-Oiwake-cho, Sakyo, Kyoto 606-8502, Japan}

\author[0000-0002-0412-0849]{Yuta Notsu}
\affil{Laboratory for Atmospheric and Space Physics, University of Colorado Boulder, 3665 Discovery Drive, Boulder, Colorado 80303, USA}

\affil{National Solar Observatory, 3665 Discovery Drive, Boulder, Colorado 80303, USA}

\author[0000-0002-1297-9485]{Kosuke Namekata}
\affil{Department of Astronomy, Kyoto University, Kitashirakawa-Oiwake-cho, Sakyo, Kyoto, Kyoto 606-8502, Japan}

\author{Taichi Kato}
\affil{Department of Astronomy, Kyoto University, Kitashirakawa-Oiwake-cho, Sakyo, Kyoto, Kyoto 606-8502, Japan}

\author[0000-0003-2493-912X]{Shota Notsu}
\altaffiliation{Current Affiliation: Star and Planet Formation Laboratory, RIKEN Cluster for Pioneering Research, 2-1 Hirosawa, Wako, Saitama 351-0198, Japan}
\affil{Leiden Observatory, Faculty of Science, Leiden University, PO Box 9513, 2300 RA Leiden, The Netherlands}

\author{Soshi Okamoto}
\affil{Department of Astronomy, Kyoto University, Kitashirakawa-Oiwake-cho, Sakyo, Kyoto, Kyoto 606-8502, Japan}

\author{Satoshi Honda}
\affil{Nishi-Harima Astronomical Observatory, University of Hyogo, 407-2, Nishigaichi, Sayo-cho, Sayo, Hyogo 679-5313, Japan}

\author{Daisaku Nogami}
\affil{Department of Astronomy, Kyoto University, Kitashirakawa-Oiwake-cho, Sakyo, Kyoto, Kyoto 606-8502, Japan}

\author{Kazunari Shibata}
\affil{Astronomical Observatory, Kyoto University, Kitashirakawa-Oiwake-cho, Sakyo, Kyoto 606-8502, Japan}

\begin{abstract} 
Starspots are thought to be regions of locally strong magnetic fields, similar to sunspots, and they can generate photometric brightness modulations. 
To deduce stellar and spot properties, such as spot emergence and decay rates, we implement computational code for starspot modeling. It is implemented with an adaptive parallel tempering algorithm and an importance sampling algorithm for parameter estimation and model selection in the Bayesian framework.
For evaluating the performance of the code, we apply it to synthetic light curves produced with 3 spots.
The light curves are specified in the spot parameters, such as the radii, intensities, latitudes, longitudes, and emergence/decay durations. The spots are circular with specified radii and intensities relative to the photosphere, and the stellar differential rotation coefficient is also included in the light curves.
As a result, stellar and spot parameters are uniquely deduced.
The number of spots is correctly determined: 
the 3-spot model is
preferable because the model evidence is much greater than that of 2-spot models by orders of magnitude and more than that of 4-spot model by a more modest factor, whereas the light curves
are produced to have 2 or 1 local minimum during one equatorial rotation period by adjusting the values of longitude.
The spot emergence and decay rates can be estimated with error less than an order of magnitude, considering the difference of the number of spots.
\end{abstract}

\section{Introduction} \label{sec:intro}
Starspots are apparent manifestations of magnetic activity on the stellar surface, and can be ubiquitously observed on various types of stars \citep[for reviews, see][]{2005LRSP....2....8B,2009A&ARv..17..251S}.
For active young stars, cool stars, and RS CVn-type stars, starspots have been extensively studied through ground-based observations \citep[e.g.,][]{1995ApJS...97..513H}.
With the advent of unprecedented precision and long-term photometry by the \textit{Kepler} space telescope \citep{2010ApJ...713L..79K}, photometric brightness modulations ascribed to spots have facilitated studies of starspot properties \citep[e.g.,][]{2010ApJ...713L.155B}.
There are some components to elucidate
stellar magnetic activities analogous to solar ones.
The most remarkable ones are the distribution of spot latitude \citep{2017ApJ...846...99M} and the degree of differential rotation \citep{2013A&A...560A...4R}.
It is also significant to measure the emergence and decay rates of starspots \citep[e.g.,][]{2019LRSP...16....3T,2019ApJ...871..187N} because
they relate to the following issues:
(i) Solar flares tend to be driven by emerging sunspots \citep{2019LRSP...16....3T}, and
superflares could occur on solar-type stars \citep{2012Natur.485..478M, 2013ApJS..209....5S, 2016NatCo...711058K, 2017PASJ...69...41M, 2019ApJ...876...58N} because large spots can cause them \citep{2013PASJ...65...49S}. (ii) Measuring magnetic diffusion of the convection zone must be constrained for theoretical stellar dynamo \citep[e.g.,][]{2014ApJ...795...79B}.
To measure the temporal evolution of starspots on solar-type stars, light curve analyses of spotted stars have been employed by utilizing the rotational brightness modulations \citep{2019ApJ...871..187N} and transiting exoplanet occultations \citep{2015PhDT.......177D,Namekata2020}.
Furthermore, light curve inversion or starspot mapping (hereinafter, referred to as starspot modeling) has also been performed to decipher starspot properties on the stellar surface \citep[e.g.,][]{1992A&A...259..183S, 2008AN....329..364S, 2009A&A...506..245M}, especially on the basis of Bayesian inference \citep{2006PASP..118.1351C, 2006ApJ...648..607C, 2007ApJ...659.1611W, 2011A&A...532A..81F, 2014A&A...564A..50L, 2018MNRAS.478..460A} using the Markov chain Monte Carlo algorithm \citep[MCMC;][]{2013PASP..125..306F, 2017ARA&A..55..213S, 2018ApJS..236...11H}.
\cite{2007ApJ...659.1611W} also implemented a parallel tempering  algorithm \citep[PT;][]{1996JPSJ...65.1604H, 2005blda.book.....G, 2005ApJ...631.1198G, 2016MNRAS.455.1919V, 2017ARA&A..55..213S} in \texttt{StarSpotz} code \citep{2006PASP..118.1351C, 2006ApJ...648..607C} to explore multi-dimensional parameter space more efficiently.

For the purpose of deducing stellar and spot properties from the photometric brightness modulations, we implement computational code for starspot modeling. It is implemented with an adaptive parallel tempering algorithm and an importance sampling algorithm for parameter estimation and model selection in the Bayesian framework \citep[cf.][]{Neal96,Neal01}. The adaptive parallel tempering algorithm is based on the PT algorithm
together with an adaptive algorithm \citep{2001Bernoulli..233H, Andrieu2008, ArakiPT2013}.
Compared with \texttt{StarSpotz} code \citep{2007ApJ...659.1611W}, we adopt the more sophisticated spotted model, including spot emergence and decay durations \citep{2012MNRAS.427.2487K}. Furthermore, the number of spots on the stellar surface can play an important role in measuring the emergence and decay rates. Several studies have investigated how the number of spots is related to light curves
 \citep[e.g.,][]{1994ApJ...420..373E, 2013ApJ...771..127N, 2017ApJ...846...99M, 2018ApJ...863..190B, 2018ApJ...865..142B,Namekata2020}.
We determine the number of spots based on model selection, computing the value of each model evidence by the importance sampling algorithm \citep[e.g.,][]{Kass1995, 2014MNRAS.441..983D}.

In this study, to evaluate the performance of the code, we revisit synthetic light curves emulating \textit{Kepler} data of spotted stars. We qualitatively evaluate how the above-described stellar and spot properties can be deduced from the photometric brightness modulations under appropriate assumptions toward conducting starspot modeling for photometric data obtained by \textit{Kepler} and 
 \textit{Transiting Exoplanet Survey Satellite} (\textit{TESS}) (Paper I\hspace{-.1em}I; Ikuta et al. 2020 in preparation).
The remainder of this paper is organized as follows. In Section \ref{sec:method}, we describe the algorithms based on Bayesian inference and a numerical setup for starspot modeling. In Section \ref{sec:result}, we discuss the results of starspot modeling in terms of parameter degeneracies, determining the number of spots, and their effects on the spot emergence and decay rates. In Section \ref{sec:summary}, we conclude this paper and describe future prospects to deal with real data.

\section{Method} \label{sec:method}
\subsection{Bayesian inference: Adaptive parallel tempering algorithm}
According to Bayes' theorem, the posterior distribution equals a product of the likelihood and the prior distribution normalized by the model evidence ${\cal Z}$:
\begin{equation} \label{Bayes}
p(\theta|{\cal D}, \textit{M})=\frac{p({\cal D}|\theta,\textit{M})p(\theta|\textit{M})}{\cal Z}, \; \; {\cal Z}=p({\cal D}|\textit{M})=\int p({\cal D}|\theta,\textit{M})p(\theta|\textit{M}) d\theta,
\end{equation}
where $\theta$, ${\cal D}$, and \textit{M} denote modeled parameters, observed data, and the assumed model, respectively.
In Bayesian inference, we compute the posterior distribution of the modeled parameters $\theta$. However, in multi-dimensional cases of more than several parameters, it becomes extremely difficult to compute the denominator of Equation (\ref{Bayes}) as the normalization constant of the posterior distribution. In such cases, we usually use the Monte Carlo method as an approximation of inference sampling.
In particular, for deducing the posterior distribution of such multi-dimensional parameters, the Markov chain Monte Carlo (MCMC) algorithm has extensively been used in astronomical context \citep{2005AJ....129.1706F,2006ApJ...642..505F, 2013PASP..125..306F, 2014ApJS..210...11N, 2017ARA&A..55..213S, 2018ApJS..236...11H}.
Especially in the case of starspot modeling of photometric data, the MCMC algorithm has been applied to photometic data obtained by \textit{Microvariability and Oscillations of Stars} (\textit{MOST}) \citep[\texttt{StarSpotz};][]{2006PASP..118.1351C, 2006ApJ...648..607C, 2007ApJ...659.1611W} and \textit{Kepler} data \citep{2011A&A...532A..81F, 2012A&A...543A.146F, 2014A&A...564A..50L, 2018MNRAS.478..460A}. 
The MCMC algorithm can generate samples that follow the posterior distribution with a proposal distribution. 
However, in the case of a multi-modal and multi-dimensional posterior distribution, the produced MCMC samples can be trapped in local maxima for so many iterations. Thus, 
the parallel tempering (PT) algorithm is occasionally implemented to circumvent this problem as in \texttt{StarSpotz} code \citep{2007ApJ...659.1611W}.
The PT algorithm introduces auxiliary distributions with a tempering parameter $\beta_l$ :
$\pi_l(\theta)\equiv \{p({\cal D}|\theta)\}^{\beta_l}p(\theta)$ (1= $\beta_1 > \cdots > \beta_l > \cdots  > 0$), where $\pi_1(\theta)$ corresponds to the posterior distribution $p(\theta|{\cal D})$ except the normalization constant.
Hereby, $\beta_l$ tempers the multi-modality of the likelihood $p({\cal D}|\theta)$, and the peaks become less pronounced as $\beta_l\rightarrow0$. It becomes easier for the corresponding Markov chain to step away from a local maximum, and chains with smaller $\beta_l$ are more readily able to explore the full parameter space. 

The PT algorithm generates multiple MCMC samples from the posterior distribution and the auxiliary distributions in parallel, and exchanges the samples of two chains between a pair of adjacent chains only for some steps. This tempering implementation and the exchange process enable local maxima to be circumvented.
The PT algorithm executes either transition or exchange at every iteration step with a probability $\alpha_r$ or $1-\alpha_r$, respectively. The value of $\alpha_r$ is determined by exploiting the trial runs, and we set $\alpha_r=0.10$.
At transition steps, as specified in the MCMC algorithm, for each $l$ a proposed $\theta_l$ for the next iteration is drawn from a proposal distribution that is chosen to be a normal distribution characterized by a variance-covariance matrix $\Sigma_l$. The proposed $\theta_l$ is accepted or rejected according to the Metropolis-Hastings algorithm \citep{1953JChPh..21.1087M, 1970Biomet..57.97H}.
At the exchange step, a sample $\theta_l$ is randomly selected and exchanged for $\theta_{l+1}$ with probability of 
$\min{\{(1, \pi_l(\theta_{l+1})\pi_{l+1}(\theta_l)/\pi_l(\theta_l)\pi_{l+1}(\theta_{l+1})\}}$.

The performance of the PT algorithm strongly depends on the tempering parameters specifically determined by their intervals, number, and proposal parameters $\Sigma_l$. These must be selected so that each chain converges as fast as possible. 
They should be finely tuned by trial-and-error in test computations so far because their relation to the number of iterations until convergence has been unclear. 
Then, the adaptive algorithm for the Metropolis-Hastings algorithm is investigated in the statistical framework \citep{2001Bernoulli..233H, Andrieu2008} and applied to astronomical data \citep{Yamada19} so that the MCMC acceptance rate in multi-dimensional case approximately converges to the optimal value $=0.25$ \citep{Roberts97}. Furthermore, \cite{ArakiPT2013} investigates the adaptive algorithm for the PT algorithm so that the PT exchange rate between adjacent chains approximately converges to the optimal value $=0.25$ \citep{Roberts98}.

The adaptive algorithm adjusts the proposal parameters using past samples during iterations on the basis of the Robbins-Monro algorithm \citep{RobbinsMonro}.  
For adaptive Metropolis algorithm with adaptive scaling, normal proposal distribution $N(\theta_l, \sigma^2_l {\bf \Sigma}_l)$ is employed as a scale factor $\sigma^2_l$
is factored out from ${\bf \Sigma}_l$, and ${\bf \Sigma}_l$ are rescaled \citep{Andrieu2008}.
At the $n$-th transition step, the $l$-th proposal parameters are updated as 
\begin{align}
{\bf \mu}_{l,n+1} &\leftarrow {\bf \mu}_{l,n} + a_n (\theta_{l,n}-{\bf \mu}_{l,n}) \\
{\bf \Sigma}_{l,n+1}&\leftarrow {\bf \Sigma}_{l,n}+ a_n \bigl((\theta_{l,n}-{\bf \mu}_{l,n}){(\theta_{l,n}-{\bf \mu}_{l,n})}^{\rm T} - {\bf \Sigma}_{l,n}\bigr) \\
{\sigma^2}_{l,n+1} &\leftarrow {\sigma^2}_{l,n} + a_n (FA_{n}-\alpha_{\textrm{\footnotesize ac}}), 
\end{align}
where $\mu_l$ is an auxiliary proposal parameter ( 
expectation value of $\theta_l$), and also updated $\mu_{l,n+1}\leftarrow \theta_{l,n+1}$ when $\theta_{l,n}$ is 
updated to $\theta_{l,n+1}$ by being exchanged for $\theta_{l-1,n}$ or $\theta_{l+1,n}$ at the exchange step.
$FA_n$ is one if the proposed sample is accepted or zero if it is rejected. $\alpha_{\textrm{\footnotesize ac}}=0.25$ is the optimal acceptance rate, which the MCMC acceptance rate should approaches 
with the proceeding of iteration \citep{Roberts97}. 
At the $n$-th exchange step, $l$-th tempering parameter is updated as 
\begin{equation}
\log \beta_{l,n+1}\leftarrow \log \beta_{l,n} -b_n (ER_{l,n}-\alpha_{\textrm{\footnotesize ex}}).
\end{equation}
$ER_{l,n}$ is one if parameters are exchanged or zero if not.
$\alpha_{\textrm{\footnotesize ex}}=0.25$ is the optimal exchange rate, which the PT exchange rate approaches 
with the proceeding of iteration \citep{Roberts98}. 
The learning coefficients $a_n$ and $b_n$ converge to zero when the number of iterations $n$ approaches infinity. The details about the choice of the learning coefficients are described in \cite{Andrieu2008}.
In this study, we determined the number of iterations $N = 4\times10^6$ after the burn-in period $=1\times10^6$ on the basis of the Gelman-Rubin convergence diagnostic \citep{1992StaSc...7..457G, Brooks98}.
We set the learning coefficients $a_n=1/(10n+N)$ and $b_n=1/(n+N)$ so that the acceptance and exchange rates adequately converge to the moderate values, and the adaptive algorithms are executed after the burn-in period of the MCMC. 
In addition, we selected the number of parallelization $l=10$ and the tempering parameters $\beta_l= \exp\{ {7(l-1)/2\}}$ by exploiting the trial runs so that chains with small $\beta_l$ are much easily able to transition in the full parameter space \citep[cf.,][]{2016MNRAS.455.1919V}.

\subsection{Importance sampling algorithm} 
For the purpose of model selection as determining the number of parameters, we compute model evidence (the denominator of Equation \ref{Bayes}) using importance sampling algorithm along with parallel tempering transition.
In Section \ref{subsec:evidence}, we compare the number of spots by the model evidence for 3-spots model, 2-spots models, and 4-spot model.
We briefly introduce the importance sampling algorithm \citep{Kass1995}.

Model evidence is approximated by Monte Carlo integration with $N$ samples as
\begin {align}
         {\cal Z} &= \int p({\cal D}|\theta,M) p(\theta|M) d\theta \simeq \frac{1}{N}\Sigma_{n=1}^N p({\cal D}|\theta_n,M), \\
         &\; \; \; \textrm{where} \; \;   \theta_n  \; \;  \textrm{is drawn from}  \; \; p(\theta|M).
    \end {align}

However, computation of the summation becomes quite inefficient if most of the likelihood $p({\cal D}|\theta,M)$ have small values and the posterior distribution $p(\theta|{\cal D},M)$ concentrates on a small region of the parameter space. ${\cal Z}$ is dominated by a few large values of the likelihood.
 
To improve the precision of the Monte Carlo integration, the above formulation is deformed with importance sampling function $q(\theta|M)$ as
\begin {align}
        {\cal Z} &= \int p({\cal D}|\theta,M) \frac{p(\theta|M)}{q(\theta|M)} q(\theta|M) d\theta \simeq \frac{ \Sigma_{n=1}^N w_n p({\cal D}|\theta_n,M)}{\Sigma_{n=1}^N w_n}, \\
        &\; \; \; \textrm{where}  \; \; w_n=\frac{p(\theta_n|M)}{q(\theta_n|M)} \; \; \; \textrm{and} \; \; \; \theta_n  \; \;  \textrm{is drawn from}  \; \; q(\theta|M).
    \end {align}

Adopting the posterior distribution $p(\theta|{\cal D},M)$ as $q(\theta|M)$ since the samples can be drawn from the posterior distribution along with the PT transition, the model evidence ${\cal Z}$ is approximated as 
\begin{equation}
 {\cal Z} \simeq \left\{ \frac{1}{N} \Sigma_{n=1}^N p({\cal D}|\theta_n,M)^{-1} \right \}^{-1}.
\end{equation}
This value converges to the precise value of the model evidence $p({\cal D}|M)$ when the number of samples $N$ approaches infinity.
Practically, we compute this value along with parallel tempering transition and use it for the model selection.

\subsection{Analytical spotted model} \label{sec:spotted}

According to \cite{2012MNRAS.427.2487K}, a spotted flux at time $t$ is described by 
the functions $A$ and $\zeta_{\pm}$ of two parameters: the angular radius of the circular spot on the surface of the star as seen from the center of the star $\alpha$ and 
the angle between the line of sight and the line from the center of the star to the spot center $\beta$. The description summed up for the number of spots $N_{\rm spot}$ is formulated as 
\begin{equation}
F({\boldsymbol \alpha},{ \boldsymbol \beta})= 1- \sum_{j=0}^4 \Biggl(\frac{jc_j}{j+4}\Biggr) - \sum_{k=1}^{N_{\rm spot}} \frac{A_k}{\pi} \Biggl[\Biggl( \sum_{j=0}^{4} \frac{4(c_j-d_j f_{{\rm spot}})}{j+4} \frac{\zeta_{+, k}^{(j+4)/2}-\zeta_{-,k}^{(j+4)/2}}{\zeta_{+,k}^2 - \zeta_{-,k}^2 } \Biggr) \Biggr],
\end{equation}  
where
\[
  A_{k} = \left\{ \begin{array}{ll}
    \pi \sin^2\alpha_{k} \cos \beta_{k} & (0< \beta_{k} < \pi/2 -\alpha_{k}) \\
    \cos^{-1}[\cos \alpha_{k} \csc \beta_{k}]  +\cos \beta_{k} \sin^2 \alpha_{k} \cos^{-1}[-\cot \alpha_{k} \cot \beta_{k}] & \\ 
    \hspace{6cm} - \cos \alpha_{k} \sin \beta_{k} \sqrt{1-\cos^2 \alpha_{k} \csc^2 \beta_{k}} & (\pi/2 -\alpha_{k} < \beta_{k} < \pi/2 +\alpha_{k}) \\
    0 & (\pi/2 +\alpha_{k} < \beta_{k} < \pi)
  \end{array} \right.
\]

\[
  \zeta_{+,k} = \left\{ \begin{array}{ll}
    \cos (\beta_{k}+\alpha_{k}) & (0< \beta_{k} < \pi/2 -\alpha_{k}) \\
    0 & (\pi/2 -\alpha_{k} < \beta_{k} < \pi)
  \end{array} \right.
\]

\[
  \zeta_{-,k} = \left\{ \begin{array}{ll}
    1 & (0< \beta_{k} < \alpha_{k}) \\
    \cos(\beta_{k}-\alpha_{k}) & (\alpha_{k} < \beta_{k} < \pi/2 + \alpha_{k}) \\
   0 & (\pi/2+\alpha_{k}< \beta_{k} < \pi).
  \end{array} \right.
\]

$A_k$ is the sky-projected area visible to the observer of spot $k$, and the inequalities within parentheses involving $\alpha_k$ and $\beta_k$ specify the conditions under which a spot is either fully visible on the near side of the star, partly visible, or fully invisible on the far side of the star, respectively.
$c_j$, $d_j$ $(j=0,1,2,3,4)$, and $f_{{\rm spot}}$ are the stellar limb-darkening coefficients, spot ones, and spot intensity relative to the photosphere, respectively. The temporal variation of $\alpha_{k}$ is represented by a trapezoidal function with time $t$ \citep[Figure 1 in][]{2012MNRAS.427.2487K}. Then, $\alpha_{k}$ and $\beta_{k}$ relate to each spot and stellar parameter as 

\[
  \alpha_{k} = \left\{ \begin{array}{ll}
    \alpha_{{\rm max},k}\{t-(t_{k}-{\cal L}_k/2-{\cal I}_k)\}/{\cal I}_k & (t_{k}-{\cal L}_k/2-{\cal I}_k< t < t_{k}-{\cal L}_k/2) \\
    \alpha_{{\rm max},k} & (t_{k}-{\cal L}_k/2< t< t_{k}+{\cal L}_k/2) \\
    \alpha_{{\rm max},k}\{(t_{k}+{\cal L}_k/2+{\cal E}_k)-t\}/{\cal E}_k& (t_{k}+{\cal L}_k/2<t<t_{k}+{\cal L}_k/2+{\cal E}_k)\\
    0& (t<t_{k}-{\cal L}_k/2-{\cal I}_k, t_{k}+{\cal L}_k/2+{\cal E}_k<t )
  \end{array} \right.
\]

\begin{align} 
\cos \beta_{k}&=\cos i \sin \Phi_{k}+ \sin i \cos \Phi_{k} \cos \{ \Lambda_{ k} +\frac{2\pi}{P(\Phi_{ k})}t \} \\
 &\; \; \; \textrm{where} \; \; \; P(\Phi_k)=\frac{P_{\rm eq}}{1-\kappa \sin^2 \Phi_k} \label{eq:diff}, 
\end{align}

where $\alpha_{{\rm max},k}$, $t_{k}$, ${\cal I}_k$, ${\cal E}_k$, and ${\cal L}_k$ are maximum radius, reference time (the time at the midpoint of the interval over which the spot has its maximum radius), emergence duration, decay duration, and stable duration, respectively. 
Each spot latitude $\Phi_{k}$ is assumed to be invariable, and each longitude is assumed to vary with time from the initial longitude $\Lambda_{k}$.
The rotation period at the latitude $\Phi_k$ is characterized by equatorial period $P_{\rm eq}$ and the degree of differential rotation $\kappa$ as solar-like differential rotation.
The limb-darkening law is adopted as a quadratic term:
\begin{equation}
I(\mu)/I(1)= 1 - u_1(1-\mu) - u_2(1-\mu)^2,
\end{equation}
where $\mu$ represents the cosine of the azimuthal angle. Then, the limb-darkening coefficients are set as $c_1=c_3=d_1=d_3=0, c_2=d_2=u_1+2u_2, c_4=d_4=-u_2$, and $c_0=d_0=1-u_1-u_2$.
We adopt solar values of the limb-darkening coefficients $c_2=d_2=0.93, c_4=d_4=-0.23$ \citep{2000asqu.book.....C}. When applied to \textit{Kepler} data of a spotted star, the limb-darkening coefficients can be adopted, dependent on the stellar parameters \citep{2010A&A...510A..21S}.

\subsection{Numerical Setup}
We employ a normal likelihood function as
\begin{align}
p ({\cal D}|\theta) &= \prod_{i} \frac{1}{\sqrt{2\pi \sigma_i^2}} \exp{\biggl[ 
-\frac{\bigl(F_{{\rm obs},i}-F_{{\rm mod},i}({\theta})\bigr)^2}{2 \sigma_i^2}
\biggr]} \\
\; \; \; \textrm{and} \; \; \; F_{{\rm mod}, i} (\theta)&=  F({\boldsymbol \alpha }_i , {\boldsymbol \beta}_i )/F_{\rm ave}-1, \label{eq:relative}
\end{align}  
where $\sigma_i$, $F_{{\rm obs},i}$, $F_{{\rm mod},i}(\theta)$, and $F_{\rm ave}$ are photometric error, relative flux scaled as Equation \ref{eq:relative} for synthetic data emulated as observation, relative model flux characterized by parameters $\theta$ at the time $t_i$, and the average of $F({\boldsymbol \alpha }_i, {\boldsymbol \beta}_i)$, respectively. 

In Table \ref{tb:para1} and \ref{tb:para2}, deduced parameters $\theta$ are denoted as the stellar and spot parameters: sine of inclination angle $\sin i$; equatorial rotation period $P_{\rm eq}$ (day); degree of differential rotation $\kappa$; relative intensity $f_{\rm spot}$; latitude $\Lambda_k$ (deg); longitude $\Phi_k$ (deg);  
reference time $t_{k}$ (day); maximum radius $\alpha_{{\rm max},k}$ (deg); emergence duration ${\cal I}_k$ (day); decay duration ${\cal E}_k$ (day); and stable duration ${\cal L}_k$ (day).
As each prior distribution, we selected truncated uniform, $\log$-uniform (Jeffery's prior), and normal distributions; it is shown that there are degeneracies between the inclination angle $i$ and each spot latitude $\Phi$, and between the spot intensity $f_{\textrm{\scriptsize spot}}$ and each spot size $\alpha$ \citep{1996ApJ...473..388E, 2013ApJS..205...17W}.
Each spot is discerned by the range of the reference time $t_k$ so that the spots are not replaced one by one during the parallel tempering parameter transition as denoted in Table \ref{tb:para1} and \ref{tb:para2}.
Furthermore, in the case of $\sin i \sim 1$, there are degeneracies between each spot latitude $\Phi$ and each spot size $\alpha$ \citep[cf. Figure 1][]{Namekata2020}. In other words, similar light curves can be generated by either a large spot at high latitude or a small spot at low latitude. Whether each spot exists on northern or southern hemisphere is also indiscernible. Therefore, we constrain the inclination angle and spot intensity as truncated normal distributions under the currently achieved precision \citep{2017PASJ...69...41M, 2019ApJ...876...58N}.

\subsection{Synthetic light curves toward \textit{Kepler} data}
Hereafter, for the purpose of modeling \textit{Kepler} data of spotted stars, we produced two synthetic light curves with 3 spots for approximately two \textit{Kepler} quarters ($\sim 200$ days) so that they have quasi-periodic modulations ascribed to the spots. These light curves are generated with the spotted model \citep{2012MNRAS.427.2487K} in addition to random error ($\sim$ 10\% of modulation amplitude), emulating most faint \textit{Kepler} stars and assuming spot-dominated stars \citep{2017ApJ...851..116M}.
The input values to produce the light curves are listed in Tables \ref{tb:para1} and \ref{tb:para2}.
Assuming inclination angle of a star is randomly distributed in real data, the expectation value equals $1$ (rad) \citep{2008oasp.book.....G}.
Then, the values of the inclination angle are set $60$ (deg), and each value of the spot latitudes are determined to be less than the value of the inclination angle so that the spots can be visible and invisible by the stellar rotation.
We note that it is difficult to deduce the parameters of always visible spots at higher latitudes than the inclination angle.
Each value of the spot longitudes is determined so that the light curves have 2 or 1 local minimum during one equatorial rotation period by adjusting the values of longitude. 
Hereinafter, we call each of the light curves
2-spot-like or 1-spot-like, respectively.
In 45-deg and 30-deg cases of the inclination angle, we also produced such light curves and optimized them as well as the 60-deg case. Thereby, we ascertained the accuracies of the deduced parameters are almost the same as that of the 60-deg case by exploiting the test runs. This is because the accuracy of the posterior distribution of the inclination angle depends on its variance of the prior distribution, and this reflects the deduced accuracies of other parameters with degeneracies with the inclination angle.
Although stellar continuum level is unknown due to some effects such as polar spots \citep[e.g.,][]{2018ApJ...865..142B}, it is assumed to be invariant for the interval of the light curves because large spots are suggested to live for a few hundred days \citep{2017MNRAS.472.1618G}.

\section{Results and Discussion} \label{sec:result}

We optimize 2-spot-like and 1-spot-like light curves by 3-spot model, 2-spot model, and 2-spot model with fixed inclination angle $\sin i$.
In each case, unimodal posterior distributions are deduced. In the 3-spot model, the modes approximately equal the input values producing the light curves.
In the 2-spot model, deduced posterior distributions have a mode of large spot radius at high latitude with higher inclination angle than the mode of the truncated normal prior distribution. Thus, the light curves are also optimized by the 2-spot model with fixed $\sin i$.
Table \ref{tb:para1} and \ref{tb:para2} show the modes of the deduced posterior distributions, their credible regions, and the model evidence for each model, together with the input values and their prior distributions for each of the parameters. 
Figure \ref{fig:plot1}, \ref{fig:plot2}, \ref{fig:plot3} and \ref{fig:plot4}, \ref{fig:plot5}, \ref{fig:plot6} show the results of the 2-spot-like case and 1-spot-like case, respectively: (a) the light curve produced with the input values of the parameters (gray), that reproduced with each mode of the deduced unimodal posterior distribution (red), and their residuals (black) and (b)
the temporal radius variation of each spot produced with the input values of the parameters (gray), and those for the 3-spot model, 2-spot model, and 2-spot model with fixed $\sin i$ (red, blue, and green), respectively. 
The inclination angle, degree of differential rotation, relative intensity, latitude, and radius have degeneracies between any of them. Their joint posterior distributions are delineated in Figure \ref{fig:corner1}, \ref{fig:corner2}, and \ref{fig:corner3} for the 2-spot-like case and Figure \ref{fig:corner4}, \ref{fig:corner5}, and \ref{fig:corner6} for the 1-spot-like case generated using open software \texttt{corner} \citep{2016JOSS....1...24F}.
The calculated spots on the stellar surface and the light curves are visualized in Figure \ref{fig:vis1} and \ref{fig:vis2}.
We discuss each degeneracy between the parameters in Section \ref{subsec:degeneracy}, model selection determining the number of spots in Section \ref{subsec:evidence}, and the effects on estimating spot lifetime in Section \ref{subsec:lifetime}.

\subsection{Degeneracy between parameters} \label{subsec:degeneracy}
\begin{description}
   \item[Inclination angle vs Spot latitude]\mbox{}\\
Inclination and each spot latitude are not uniquely deduced under uniform prior distributions due to  the degeneracies \citep{1996ApJ...473..388E, 2013ApJS..205...17W}.
Therefore, the sine of the inclination angle is constrained as a truncated normal prior distribution with the center equivalent to the input value ($= \sin 60^{\circ}$) and with the variance ($=0.15^2$) based on currently achieved precision of spectroscopy (Tables \ref{tb:para1} and \ref{tb:para2}) \citep[][]{2014PASJ...66L...4N, 2013PASJ...65..112N, 2015PASJ...67...32N, 2015PASJ...67...33N, 2019ApJ...876...58N}. 
Then, posterior distributions of the inclination angle and spot latitudes are unimodally and adequately deduced from the photometric light curve. We note that the posterior distributions of the inclination angle deduced from the light curve is likely to have a higher accuracy than that would be deduced from real data.

   \item[Spot relative intensity vs Spot radius]\mbox{}\\
Spot relative intensity and each spot size are not uniquely deduced 
under the uniform prior distributions due to  the degeneracies \citep{2013ApJS..205...17W}.
Therefore, the spot relative intensity is constrained as a normal prior distribution with the center equivalent to the input value ($= 0.30$) and with the variance ($=0.05^2$) adopted from
a formula of the spot 
temperature based on the Doppler imaging technique (Tables \ref{tb:para1} and \ref{tb:para2}) \citep[e.g.,][]{2005LRSP....2....8B}. Then, the spot relative intensity and each spot radius are unimodally and adequately deduced from the photometric light curve.  
We note that the posterior distributions of the relative intensity deduced from the light curve is also likely to have a higher accuracy than that would be deduced from real data.

   \item[Differential rotation vs Spot latitude]\mbox{}\
There are degeneracies between the degree of differential rotation and the spot latitudes due to adjusting the periodicity for each spot (Equation \ref{eq:diff}).  
However, the deduction of the spot latitude depends on that of the inclination angle, and thus the degree of differential rotation is unimodally deduced from the photometric light curve only if the number of spots is more than two.

   \item[Spot latitude vs Spot radius]\mbox{}\\
There are degeneracies between spot latitudes and the spot radii because the same modulation amplitude is generated by adjusting the parameters 
\citep[cf.][]{Namekata2020}. However, the deduction of the spot latitude depends on that of the inclination angle, and thus the spot radius is unimodally deduced from the photometric light curve. We note that when the inclination angle becomes too small, the spots are always visible and do not significantly modulate the light curve.
\end{description}

\subsection{Model selection: How many spots exist?} \label{subsec:evidence}
More spots are observationally indicated to exist than seen in the light curve \citep{2017ApJ...846...99M, Namekata2020}, whereas the light curve produced with many spots is similar to that with 2 spots or 1 spot \citep{1994ApJ...420..373E, 2018ApJ...865..142B}. 
Then, we determine the number of spots based on model selection in the Bayesian framework \citep{Kass1995}. We compute the model evidence using the importance sampling algorithm along with the parallel tempering transition and compare each model. 
The values of the model evidence 
$\log {\cal Z}$ are listed in the Table \ref{tb:para1} and \ref{tb:para2}. In both of cases, the 3-spot model is much more decisive than the 2-spot model and the 2-spot model with fixed $\sin i$ by orders of magnitude:
for the 2-spot-like case, the evidence of the 3-spot model relative to that of the 2-spot model and the 2-spot model with fixed $\sin i$ are $\Delta \log {\cal Z}=2639.880$ and $2750.157$, respectively.
For the 1-spot-like one, they are $\Delta \log {\cal Z}=900.968$ and $951.955$, respectively.
The difference of the values of the model evidence for the 2-spot-like light curve is much larger than that of the 1-spot-like one because the 2-spot-like one is much more informative to deduce spot properties, such as spot emergence and decay rates \citep{2019ApJ...871..187N}.
In addition, when optimizing light curves by the 4-spot model, the parallel tempering sampling converge to the a multi-modal distribution with much peaks and with degeneracies between the parameters. 
For the 2-spot-like case, the values of the model evidence of the 4-spot model and the evidence of the 3-spot model relative to that of the 4-spot model are $\log {\cal Z}=60257.257$ and $ \Delta \log {\cal Z}=0.314$, respectively. For the 1-spot-like case, they are  $\log {\cal Z}=60305.432$ and $ \Delta \log {\cal Z}=1.916$, respectively.
Then, the 3-spot model is 
preferable, and the number of spots can be correctly determined in the case of the synthetic light curve. We note that, when conducting starspot modeling of real data, spots are not completely circular, and small spots can be ignored.

\subsection{Effect on estimating spot emergence and decay rates}\label{subsec:lifetime}
The number of spots can directly affect measuring emergence and decay rates.
For instance, when optimizing the light curve produced with 3 spots by the 2-spot models, 2 spots out of 3 spots behave as 1.
Thus, we qualitatively evaluate an estimation of the spot emergence and decay rates ($\sim$ $\alpha_{{\scriptsize \textrm{max},k}}^2/{\cal I}_k$, $\alpha_{{\scriptsize \textrm{max},k}}^2/{\cal E}_k$).
Relative to the 3-spot model, the 2-spot model overestimates by a factor of up to 6 because the inclination angle is large and the spot is at high latitude. 
In the 2-spot model with fixed $\sin i$, the estimates are larger than those of the 3-spot model by a factor of up to 2.
These values can have an error of an order of magnitude in the range of the photometic error. 

\section{conclusion and future prospects} \label{sec:summary}
\begin{figure}[ht!] 

\plotone{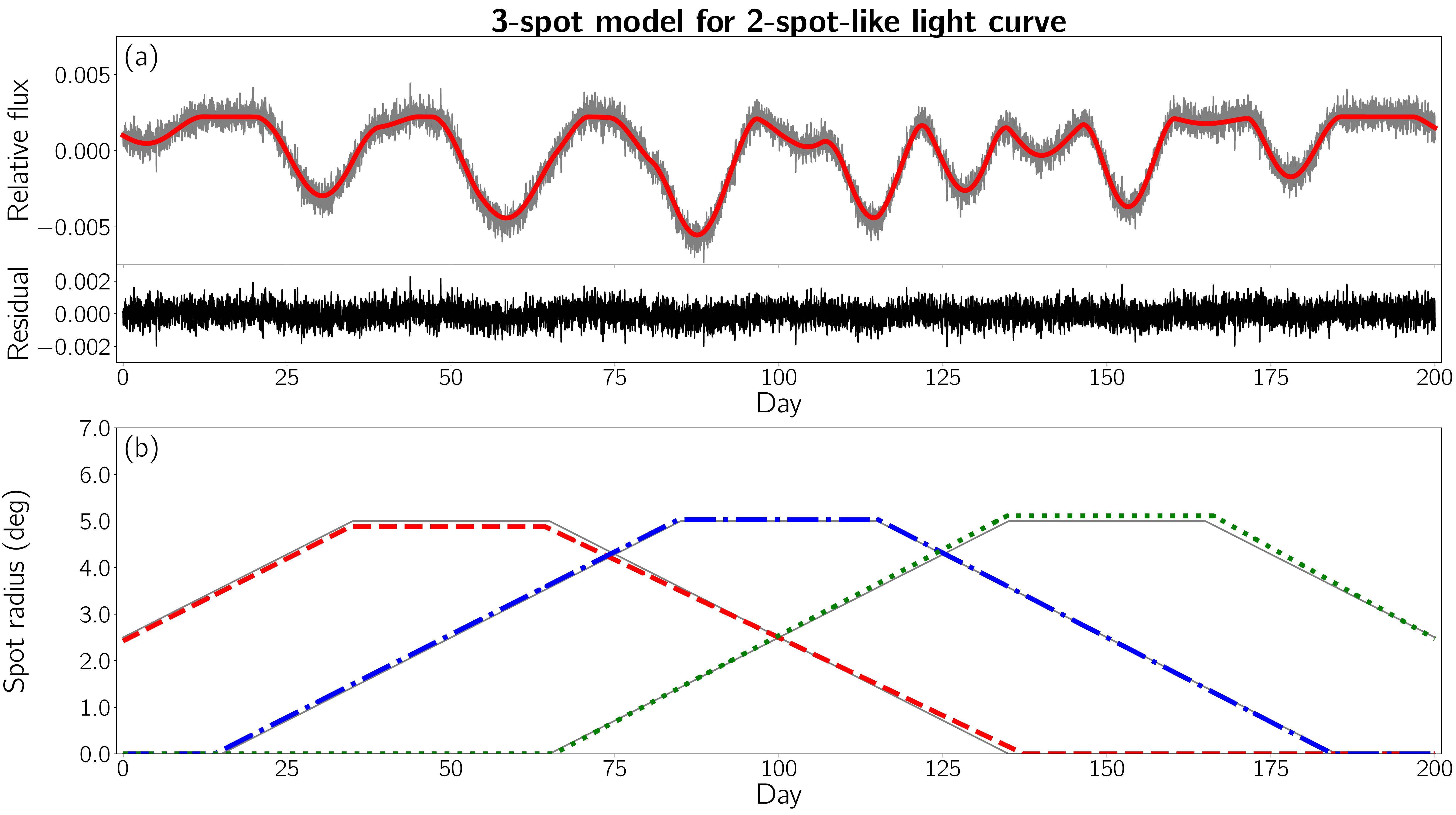}
\caption{(a) 2-spot-like light curves produced with the input values of the parameters (gray), those reproduced with each mode of the deduced unimodal posterior distribution for the 3-spot model (red), and their residuals (black); (b) Temporal radius variation of each spot produced with the input values of the parameters (gray), and that of the 3-spot model (red, blue, and green). \label{fig:plot1}}
\end{figure}

\begin{figure}[ht!] 
\plotone{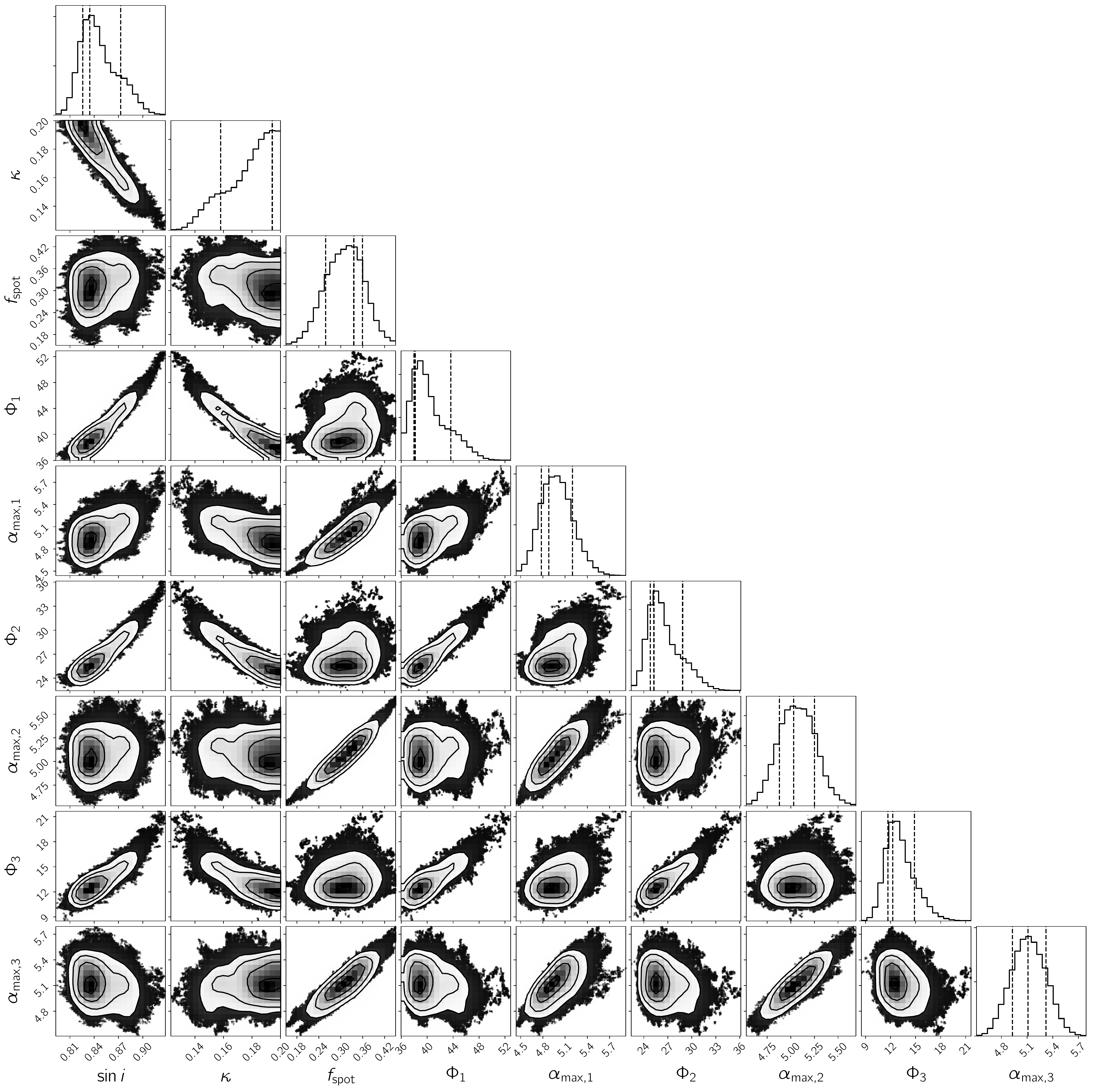}
\caption{The joint posterior distribution 
of parameters with the degeneracies for 2-spots-like light curve by the 3-spot model. Each column represents the inclination angle $\sin i$, degree of differential rotation $\kappa$, relative intensity $f_{\scriptsize \textrm{spot}}$, maximum radius $\alpha_{{\scriptsize \textrm{max}, k}}$, and latitude $\Phi_k$. \label{fig:corner1}} 
\end{figure}

\begin{figure}[ht!]
\plotone{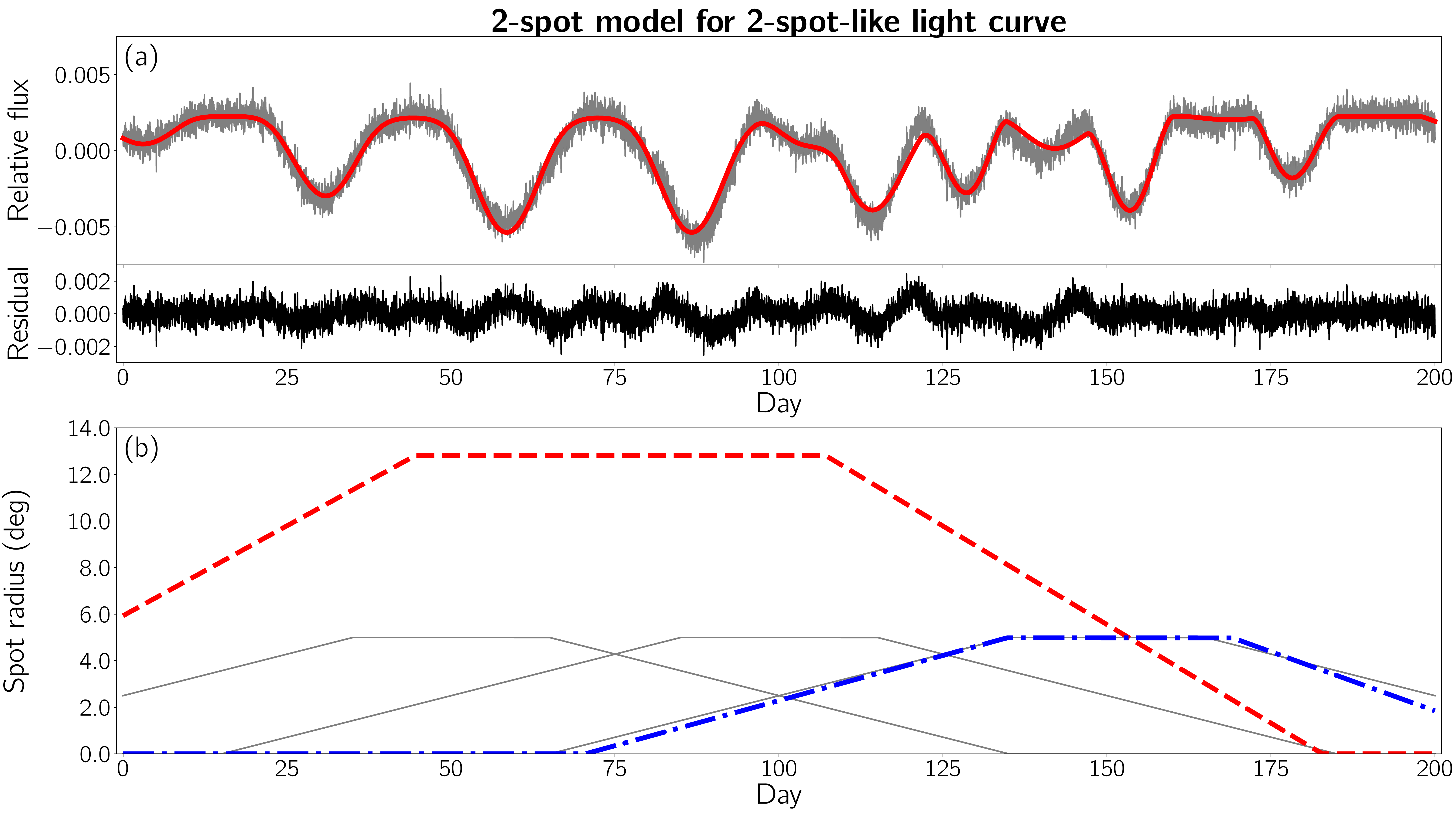}
\caption{(a,b) The same as Figure \ref{fig:plot1} but for the 2-spot model. \label{fig:plot2}}
\end{figure}

\begin{figure}[ht!]
\plotone{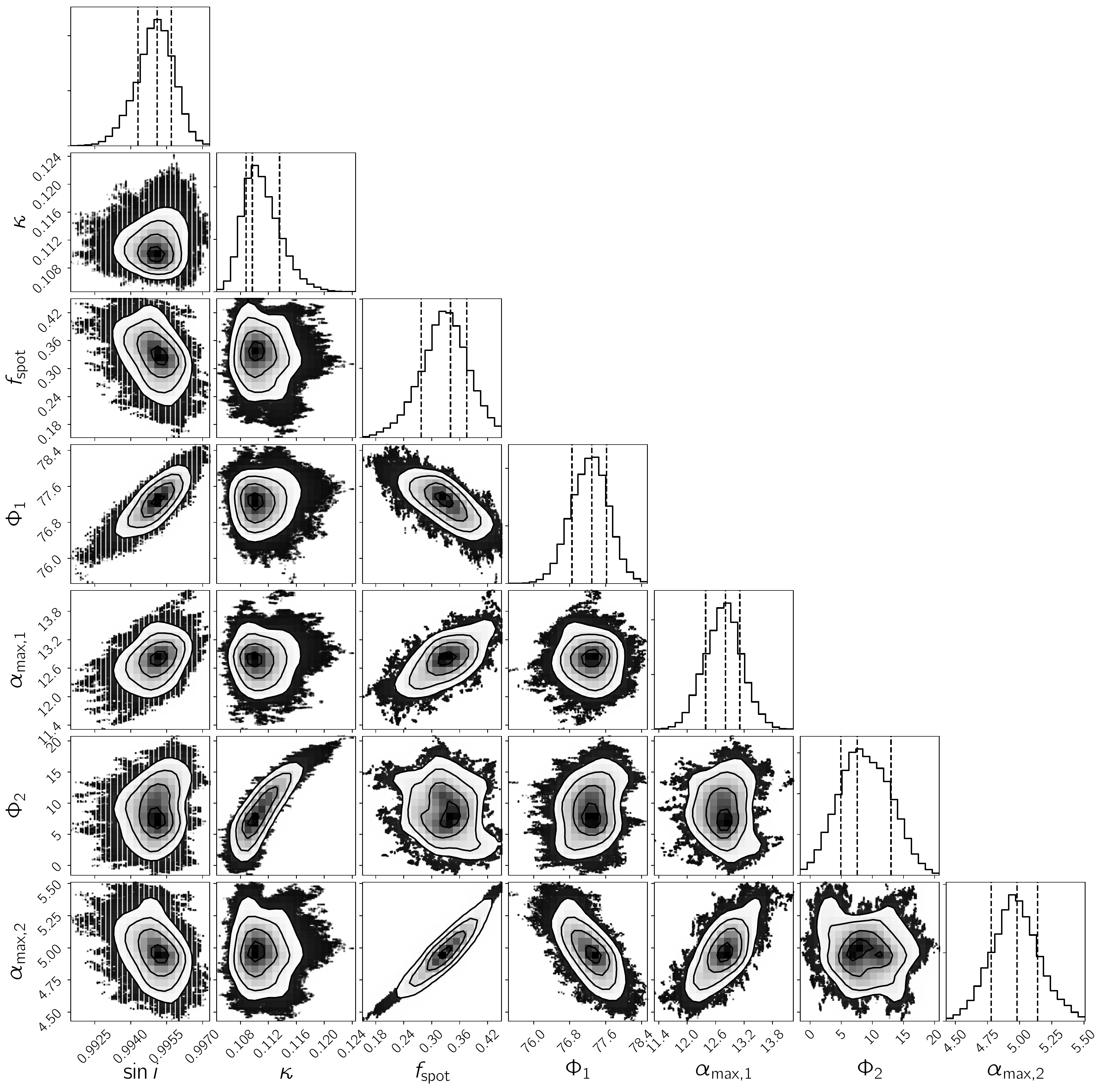}
\caption{The same as Figure \ref{fig:corner1} but for the 2-spot model. \label{fig:corner2}} 
\end{figure}

\begin{figure}[ht!]
\plotone{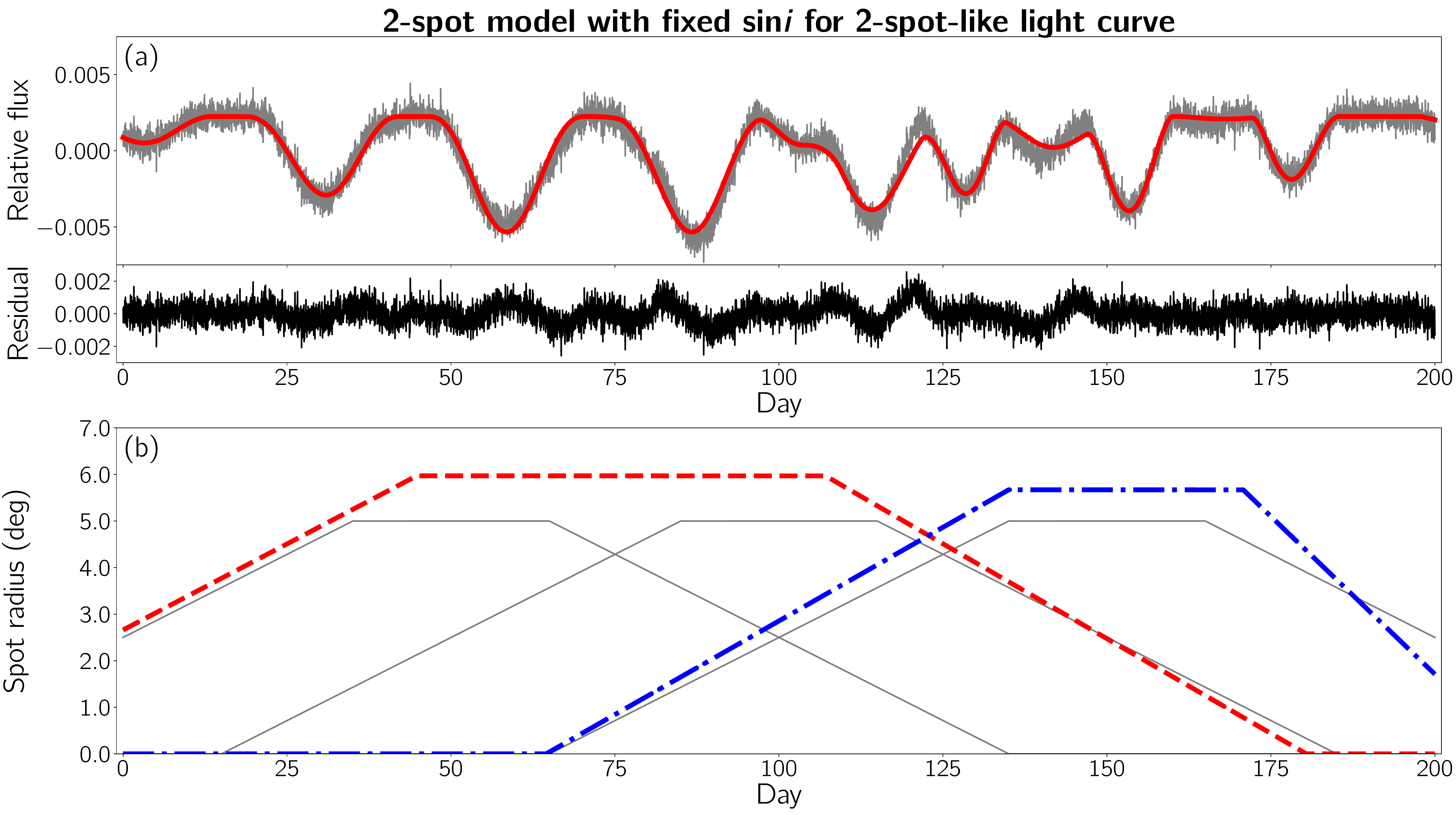}
\caption{(a,b) The same as Figure \ref{fig:plot1} but for the 2-spot model with fixed $\sin i$.
\label{fig:plot3}}
\end{figure}

\begin{figure}[ht!]
\plotone{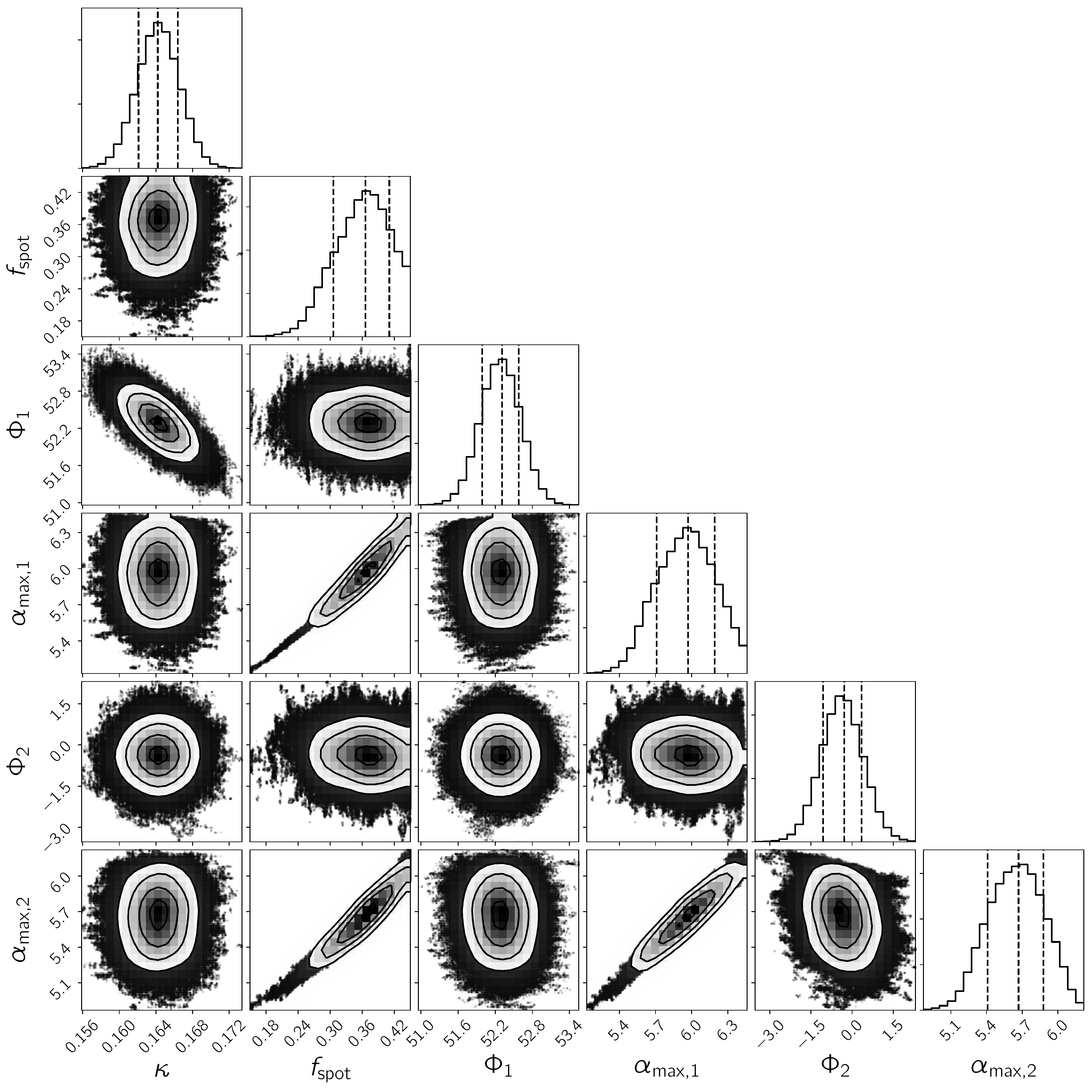}
\caption{The same as Figure \ref{fig:corner1} but for the 2-spot model with fixed $\sin i$.  \label{fig:corner3}} 
\end{figure}

\begin{figure}[ht!]
\plotone{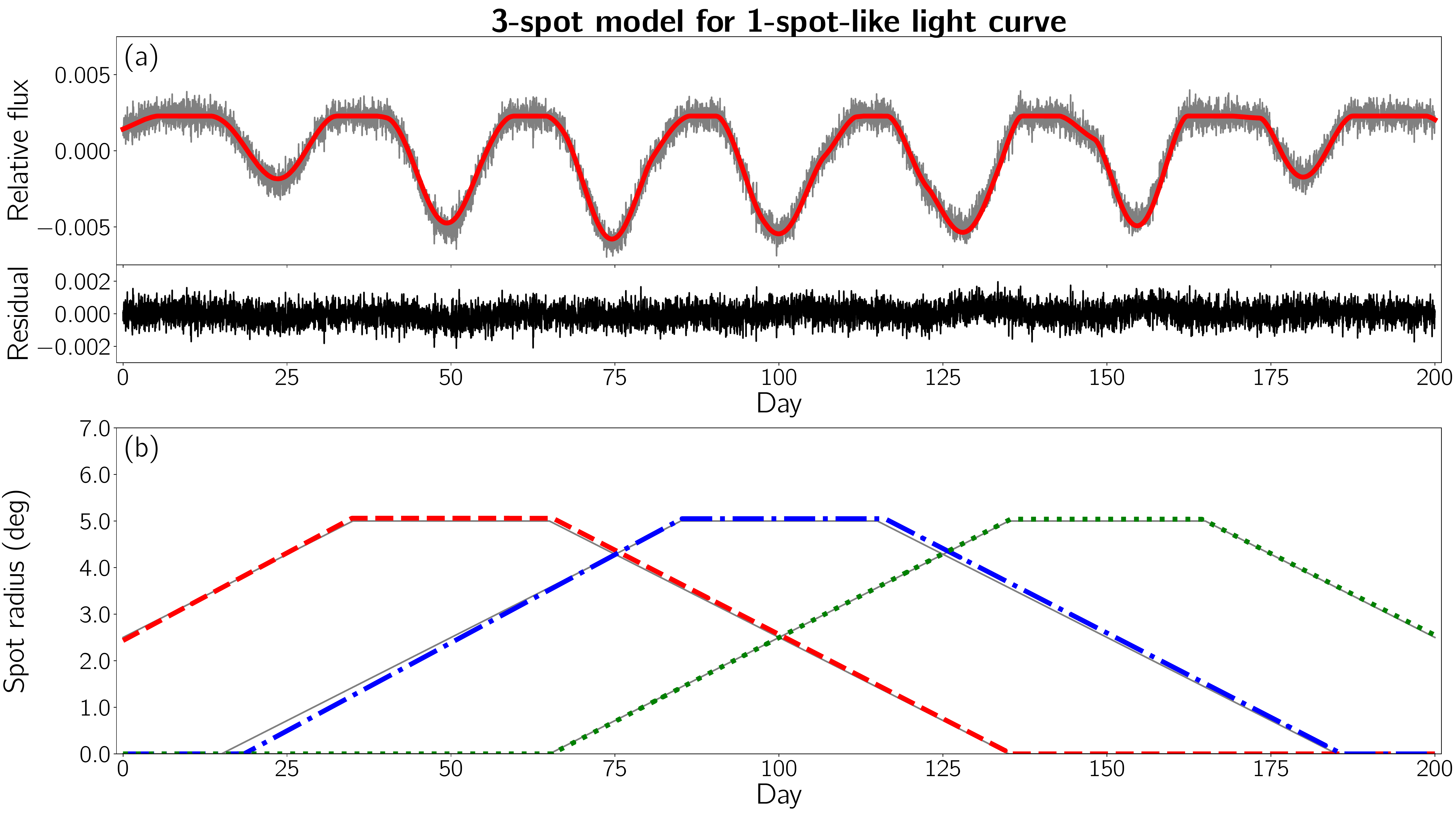}
\caption{(a) 1-spot-like light curves produced with the input values of the parameters (gray), those reproduced with each mode of the deduced unimodal posterior distribution for the 3-spot model (red), and their residuals (black); (b) Temporal radius variation of each spot produced with the input values of the parameters (gray), and that of the 3-spot model (red, blue, and green). \label{fig:plot4}}
\end{figure}

\begin{figure}[ht!]
\plotone{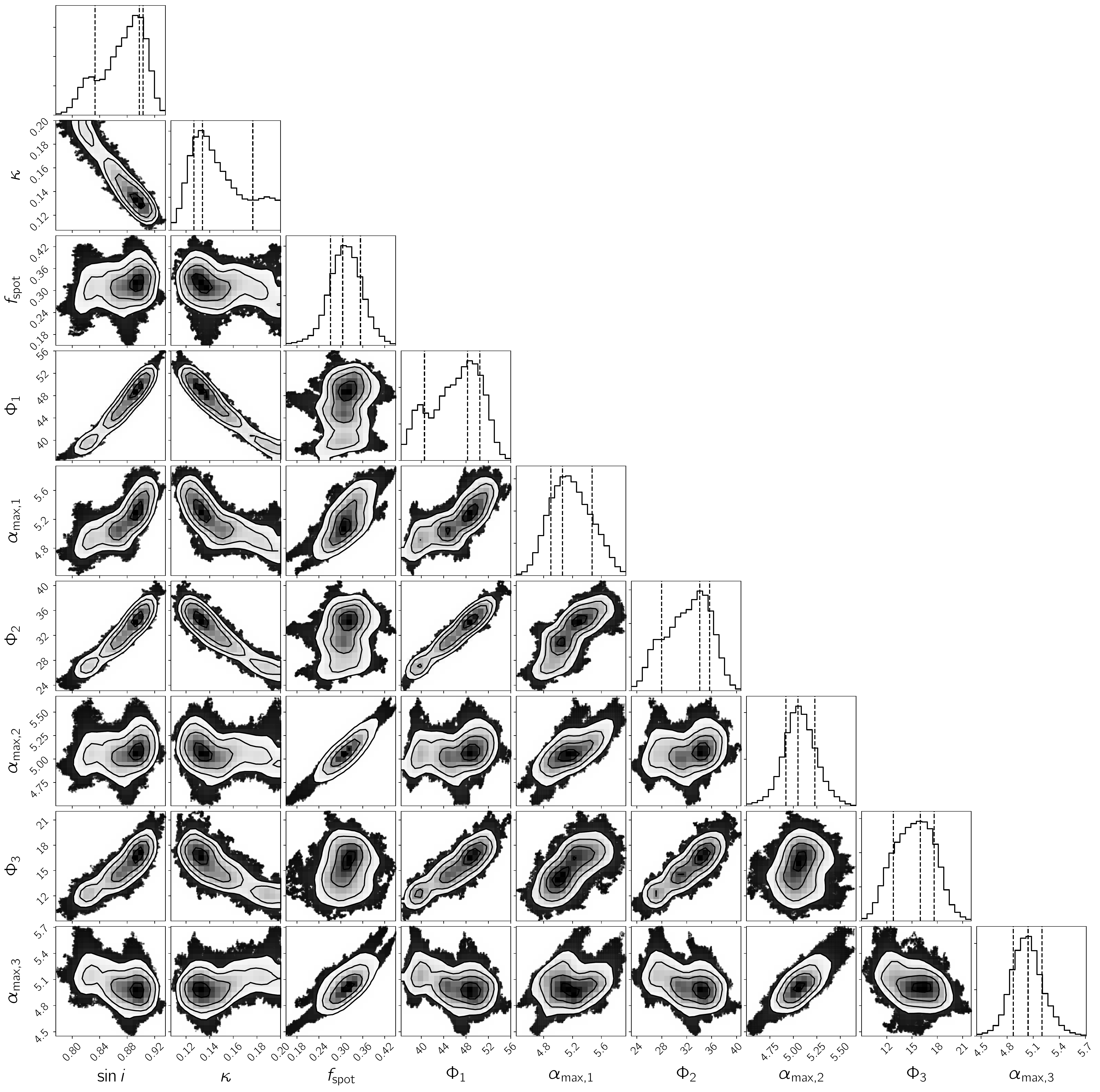}
\caption{The joint posterior distribution
of parameters with the degeneracies for 1-spots-like light curve by the 3-spot model. Each column represents the inclination angle $\sin i$, degree of differential rotation $\kappa$, relative intensity $f_{\scriptsize \textrm{spot}}$, the maximum radius $\alpha_{{\scriptsize \textrm{max,k}}}$, and latitude $\Phi_k$. \label{fig:corner4}} 
\end{figure}

\begin{figure}[ht!]
\plotone{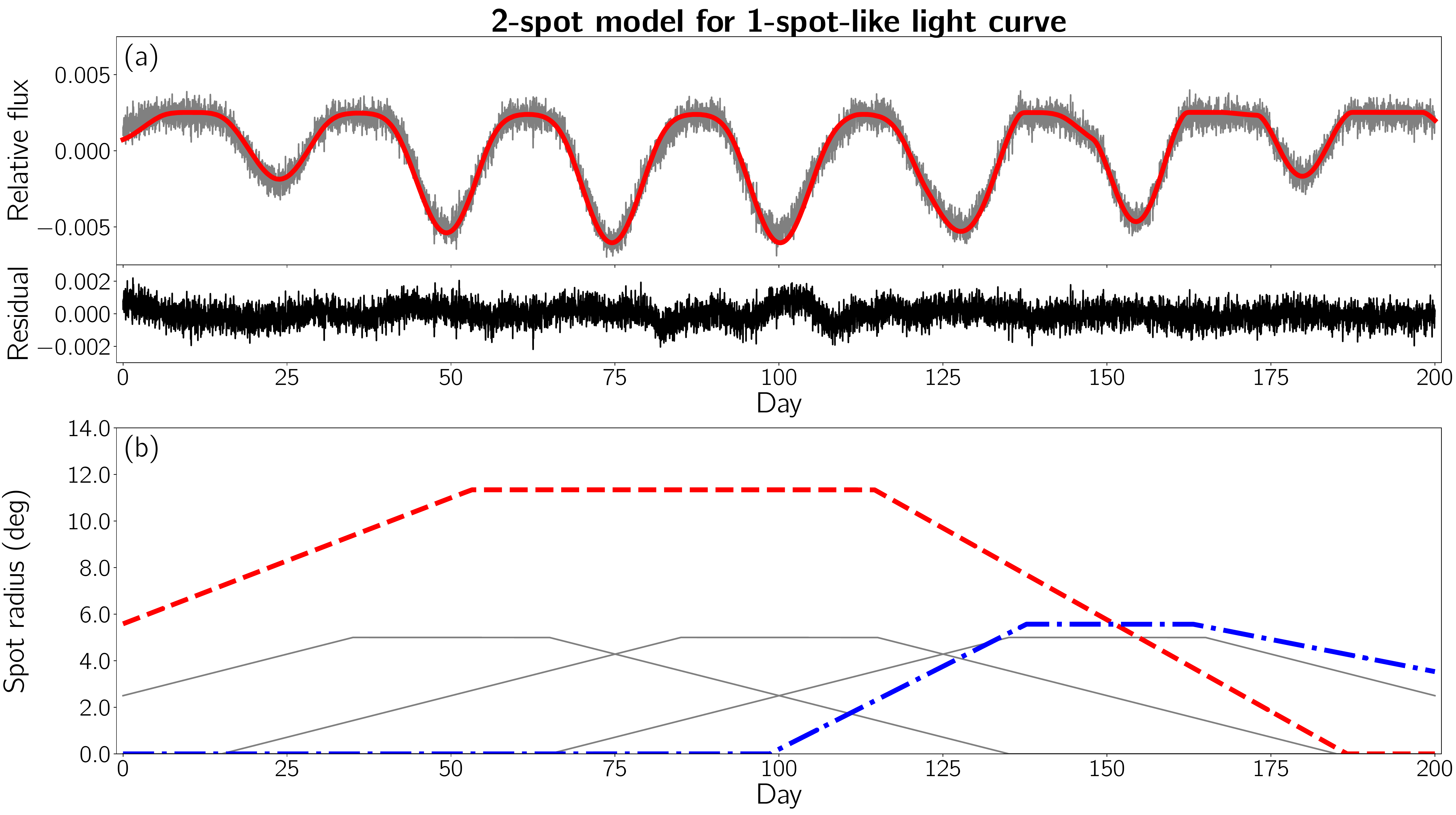}
\caption{(a,b) The same as Figure \ref{fig:plot4} but for the 2-spot model. \label{fig:plot5}}
\end{figure}

\begin{figure}[ht!]
\plotone{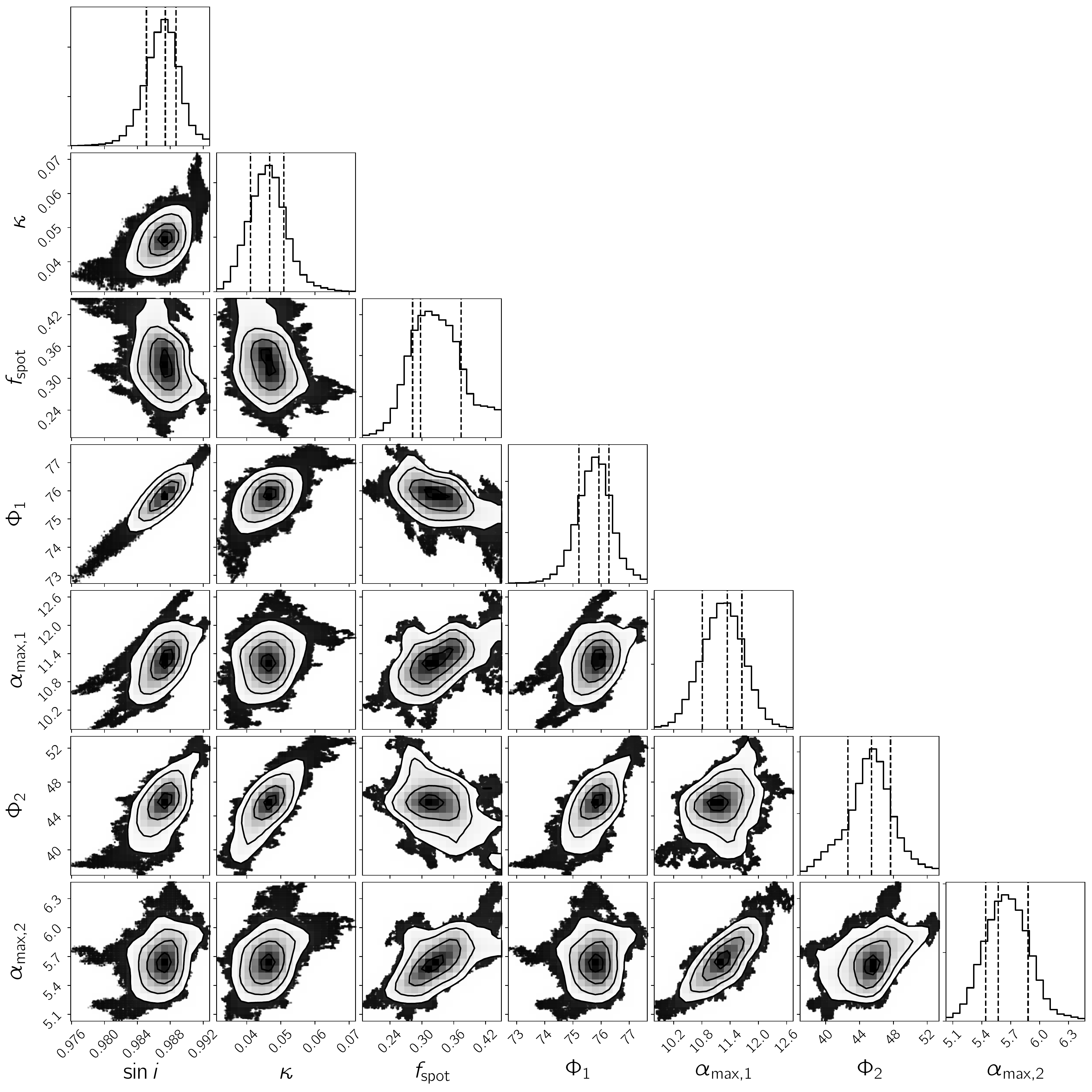}
\caption{The same as Figure \ref{fig:corner4} but for the 2-spot model. \label{fig:corner5}} 
\end{figure}

\begin{figure}[ht!]
\plotone{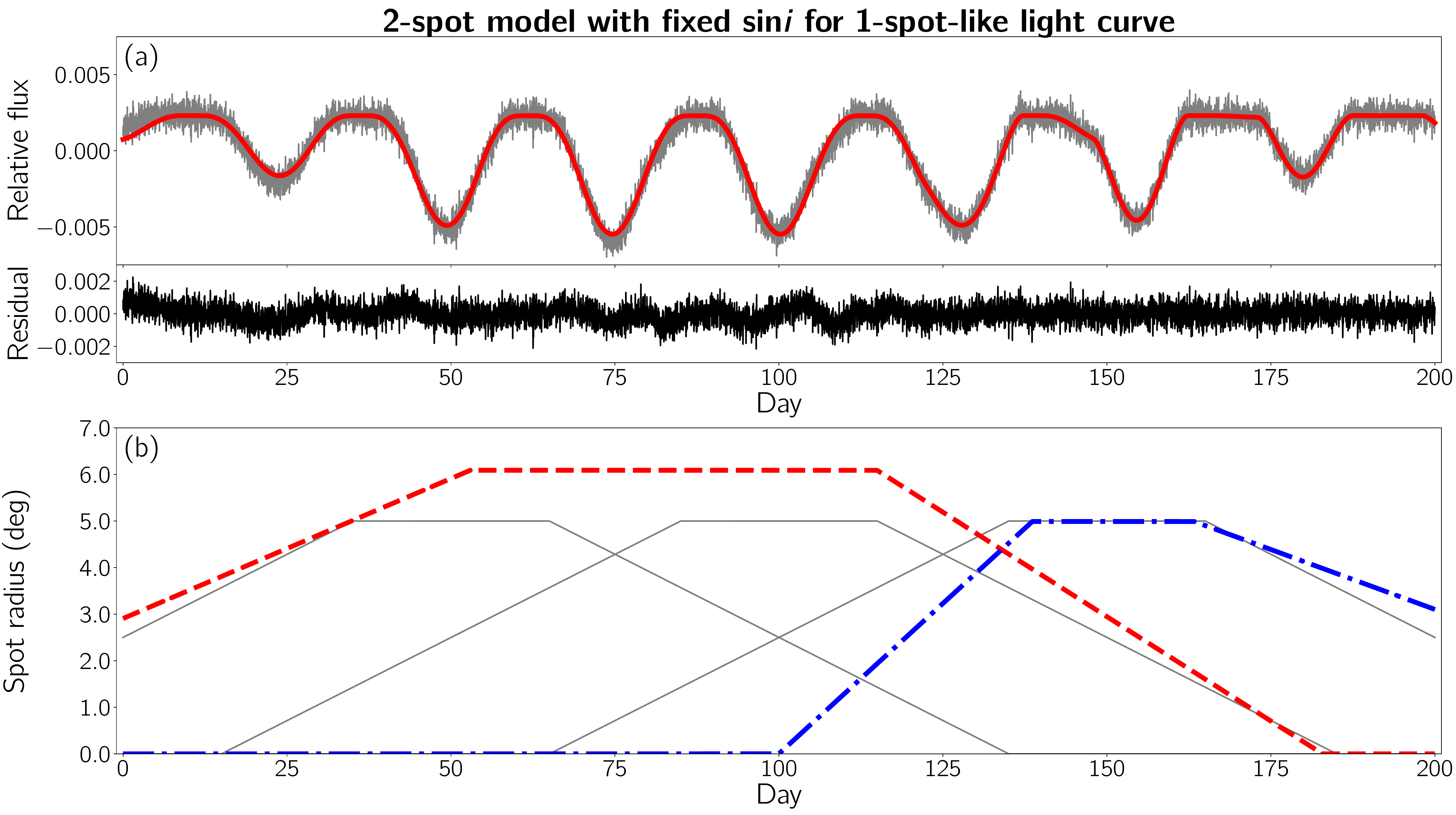}
\caption{(a,b) The same as Figure \ref{fig:plot4} but for the 2-spot model with fixed $\sin i$. \label{fig:plot6}}
\end{figure}

\begin{figure}[ht!]
\plotone{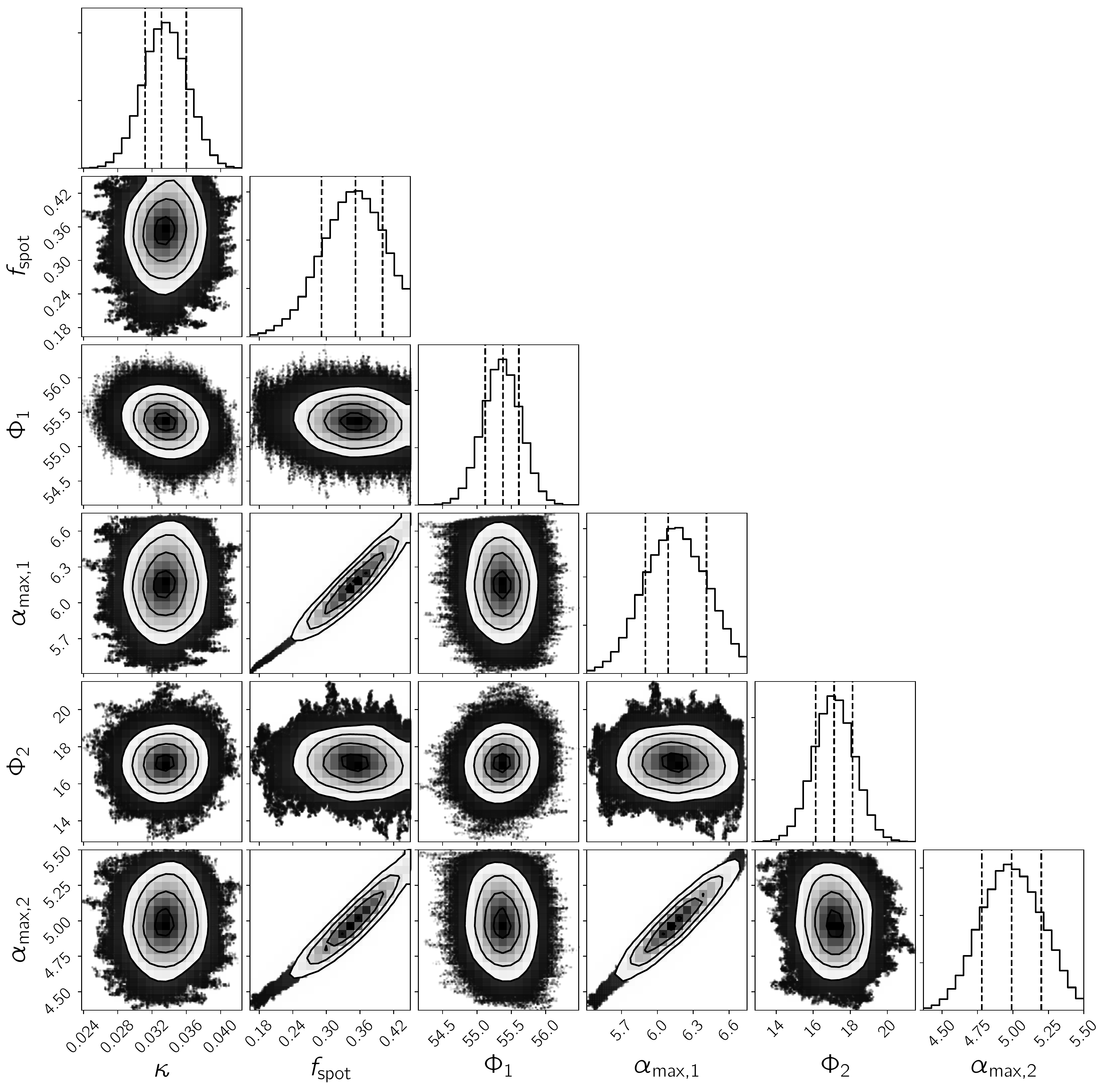}
\caption{The same as Figure \ref{fig:corner4} but for the 2-spot model with fixed $\sin i$. \label{fig:corner6}} 
\end{figure}

\begin{figure}[ht!]
\plotone{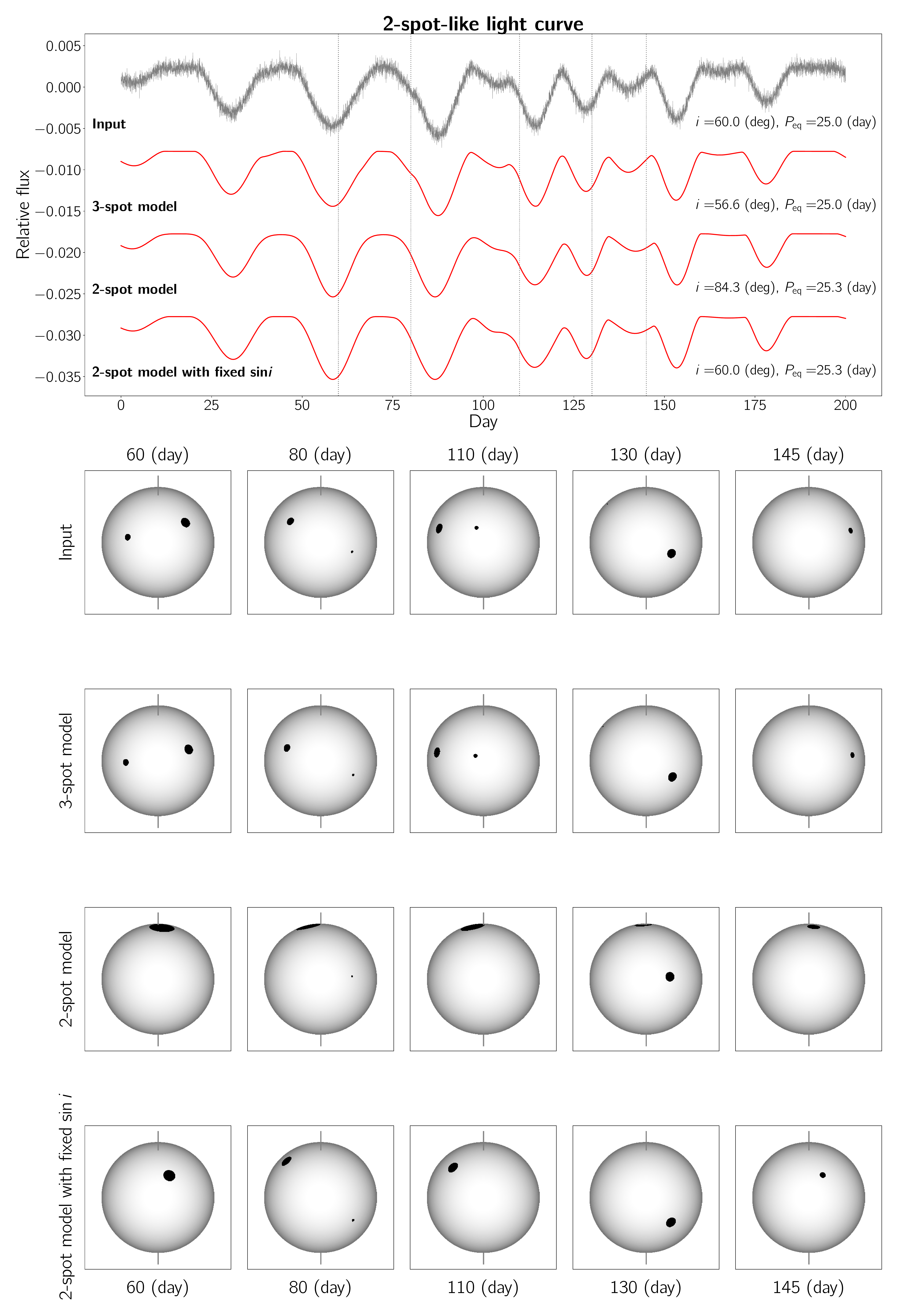}
\caption{The input light curve (gray) and the ones reproduced by the optimum of each of the models (red) for the 2-spot-like case. The values of the inclination angle and the equatorial period are also denoted for each of the models. The calculated spots on the stellar surface are visualized at five times (vertical dotted lines).\label{fig:vis1}} 
\end{figure}

\begin{figure}[ht!]
\plotone{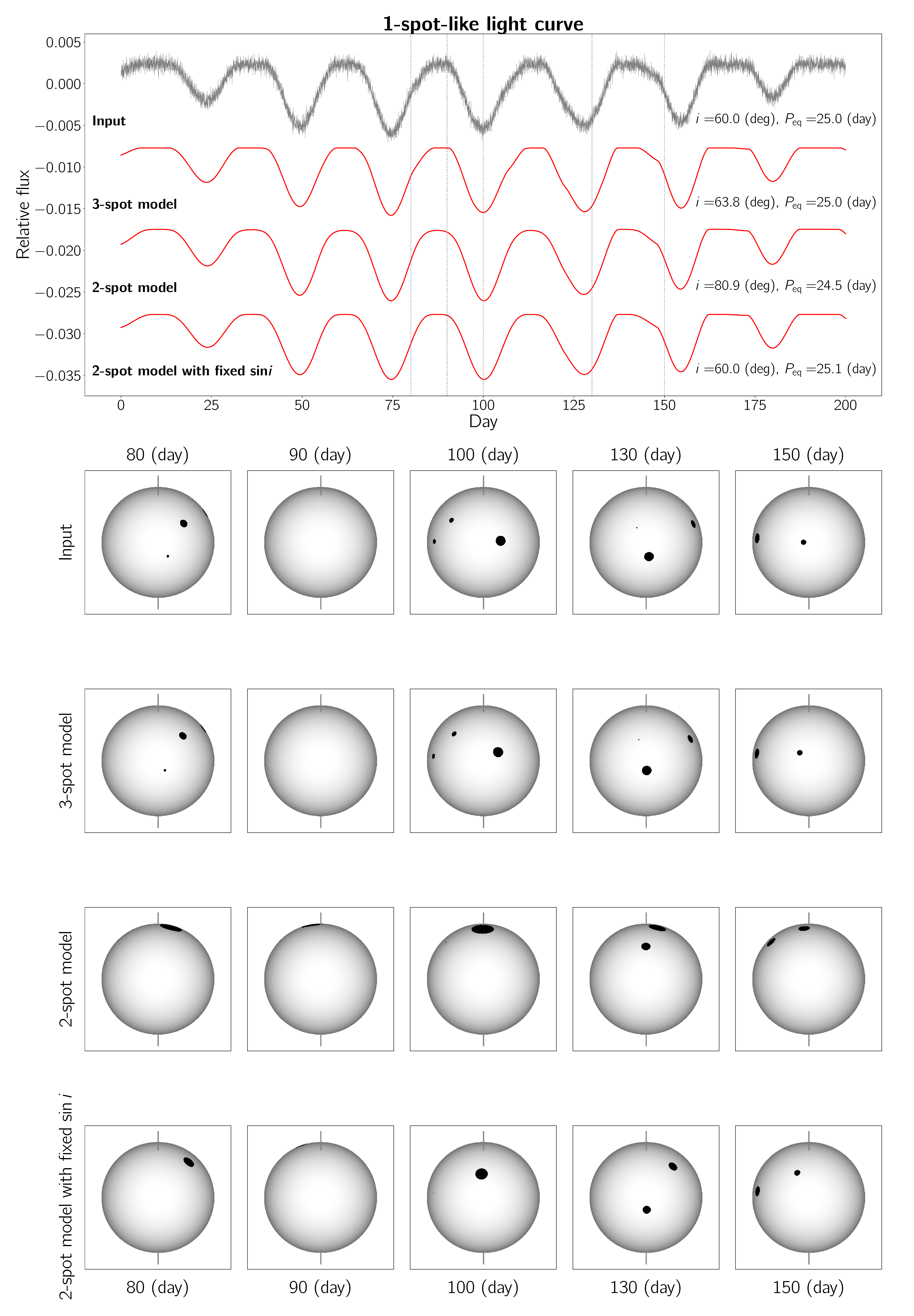}
\caption{The same as Figure \ref{fig:vis1} but for the 1-spot-like case.\label{fig:vis2}} 
\end{figure}

We implement computational code for starspot modeling to deduce stellar and spot properties from photometric brightness modulations.
It is implemented with an adaptive parallel tempering algorithm and an importance sampling algorithm for parameter estimation and model selection in the Bayesian framework. 
In this paper, for evaluating the performance of the code, we apply it to synthetic light curves emulating \textit{Kepler} data of spotted stars.
The light curves are specified in the spot parameters, such as the radii, intensities, latitudes, longitudes, and emergence/decay durations, and produced with 3 spots so that they have 2 or 1 local minimum during one equatorial rotation period by adjusting
the values of longitude. The spots are circular with specified radii and intensities relative to the photosphere, and
the stellar differential rotation coefficient is also included in the light curves.
We conduct starspot modeling for the light curves, optimizing by the 3-spot model (Figure \ref{fig:plot1}, \ref{fig:plot4}), 2-spot model (Figure  \ref{fig:plot2},  \ref{fig:plot5}), and 2-spot model with fixed $\sin i$ (Figure \ref{fig:plot3}, \ref{fig:plot6}). The calculated spots on the stellar surface and the light curves are visualized (Figure \ref{fig:vis1}, \ref{fig:vis2}).
To determine the number of spots, we compare the value of the model evidence for each model. 
In Section \ref{sec:result}, we describe the results, which can be summarized as follows:
\begin{itemize}
    \item[(i)] Unimodal posterior distributions are deduced in all of the models (Table \ref{tb:para1}, \ref{tb:para2}). In the 3-spot model, of course, the modes of the posterior distribution approximately equal the input values of the parameters producing the synthetic light curves. Then, the degeneracies between the parameters
    are eliminated by constraining the inclination angle and the relative intensity with truncated normal prior distributions 
    (Figure \ref{fig:corner1}, \ref{fig:corner2}, and \ref{fig:corner3} for 2-spot-like case and Figure \ref{fig:corner4}, \ref{fig:corner5}, and \ref{fig:corner6} for 1-spot-like case).
    \item[(ii)] The 3-spot model is decisive because the model evidence is much larger than that of the 2 spot model or 2-spot model with fixed $\sin i$ by orders of magnitude (Table \ref{tb:para1}, \ref{tb:para2}). Optimizing light curves by 4-spot model, the parallel tempering sampling converge to a multi-modal distribution with much peaks and with degeneracies between the parameters. Comparing the value of the model evidence with that of the 3-spot model, the 3-spot model is
    preferable, and the number of spots can be correctly determined in the case of the synthetic light curve.
    \item[(iii)] Spot emergence and decay rates can be estimated within an error less than an order of magnitude, considering the 3-spot model, 2-spot model, and 2-spot model with fixed $\sin i$.
\end{itemize}

In the following paper (Paper I\hspace{-.1em}I), we intend to conduct starspot modeling for \textit{Kepler} and \textit{TESS} data of spotted stars. In particular, \textit{Kepler} data include solar-type stars on which superflares are reported \citep[][Okamoto et al. 2020 in preparation]{2019ApJ...876...58N}. 
We note that \textit{Kepler} data include a long-term trend and instrumental noise, and their unspotted level is unknown \citep[e.g.,][]{2018ApJ...865..142B}. It is also necessary to determine the inclination angle precisely by another method, such as spectroscopic observation, when conducting starspot modeling. Then, we can investigate the connection between superflares and stellar and spot properties deduced by starspot modeling and compare the results of measuring emergence and decay rates with those by other methods \citep{Namekata2020}. Bright spotted stars have been observed by \textit{TESS} \citep{2014SPIE.9143E..20R}, and superflares on hundreds of spotted solar-type stars have been reported \citep{2020ApJ...890...46T,2020arXiv200507710F}.
Some \textit{TESS} targets are to be simultaneously observed by the \textit{Seimei} telescope in Kyoto University \citep{Kurita2020} using the high dispersion spectrograph. This could allow us to obtain informative prior knowledge for conducting starspot modeling of \textit{TESS} data.

\acknowledgements

K.I. sincerely appreciates Makoto Uemura for his statistical advices and Naoto Kojiguchi for his technical advices on \textit{Python} programming. 
Numerical computations were carried out on the PC cluster at Center for Computational Astrophysics, National Astronomical Observatory of Japan and a Cray
XC40 at the Yukawa Institute for Theoretical Physics, Kyoto
University. Our study was also supported by JSPS KAKENHI Grant Numbers 
JP25120007(TK), 
JP16H03955(KS), 
JP16J00320(YN), 
JP16J06887(SN), 
JP17H02865(DN), 
JP17K05400(HM), 
and JP18J20048(KN).

\begin{deluxetable*}{lccccc}
\tablecaption{2-spot-like light curve case\label{tb:para1}}
\tabletypesize{\footnotesize}
\tablehead{
\colhead{Deduced parameters} & \colhead{Input} & \colhead{3-spot model} & \colhead{2-spot model} & \colhead{2-spot model}& \colhead{Prior distribution\tablenotemark{a}}  \\
\colhead{} &\colhead{value} &\colhead{} &\colhead{} & \colhead{with fixed $\sin i$}&\colhead{} 
}
\startdata
\textbf{(Stellar parameters)} &&&&&  \\
1.  Sine of inclination angle $\sin i$& 0.8660   & $0.8346^{+0.0382}_{-0.0090}$ &$0.9951^{+0.0005}_{-0.0009}$ & 0.8660 (\textit{fixed})   &${\cal TN}(0.8660,0.1500^2,0.0000,1.0000)$\tablenotemark{b} \\
2.  Equatorial period $P_{{\scriptsize \textrm{eq}}}$ (day) & 25.0000 & $25.0431^{+0.0376}_{-0.0525}$ & $25.2645^{+0.0152}_{-0.1160}$&$25.3125^{+0.0228}_{-0.0202}$ &${\cal U}_{\log} (24.0000, 26.0000)$\\
3.   Degree of differential rotation $\kappa$ &0.1500 & $0.1941^{+0.0002}_{-0.0363}$ & $0.1097^{+0.0038}_{-0.0010}$ &$0.1642^{+0.0021}_{-0.0022}$ &${\cal U} (0.0000,0.2000)$\\
\textbf{(Spot parameters)} &  & &&&   \\
4.     Relative intensity $f_{\scriptsize \textrm{spot}}$ & 0.3000 & $0.3356^{+0.0239}_{-0.0772}$ &$0.3403^{+0.0351}_{-0.0633}$ &$0.3658^{+0.0449}_{-0.0601}$   &{${\cal TN} (0.3000,0.0500^2,0.1500,0.4500)$  \tablenotemark{c}}\\
(\textit{1st spot}) &  &  & && \\
5.     Latitude $\Phi_1$  (deg) &45.00& $38.14^{+5.52}_{-0.18}$ & $77.29^{+0.32}_{-0.45}$ &$52.31^{+0.26}_{-0.33}$   &${\cal U} (-90.00,90.00)$\\
6.     Initial longitude $\Lambda_1$  (deg)&-35.00 & $-34.30^{+0.69}_{-0.71}$  & $-26.30^{+0.39}_{-0.41}$&$-26.00^{+0.39}_{-0.41}$   &${\cal U}  (-180.00,180.00)$\\
7.     Reference time $t_{{\scriptsize 1}}$ (day) & 50.00& $49.51^{+0.29}_{-0.33}$ & $75.81^{+0.16}_{-0.20}$ &$75.93^{+0.16}_{-0.20}$    &{${\cal U} (0.00,t_{{\scriptsize 2}})$\tablenotemark{d}} \\
8.     Maximum radius $\alpha_{{\scriptsize \textrm{max,1}}}$  (deg)& 5.00& $4.88^{+0.31}_{-0.11}$ & $12.81^{+0.29}_{-0.44}$&$5.97^{+0.21}_{-0.28}$    &${\cal U} (0.01,15.00)$\\
9.     Emergence duration ${\cal I}_1$ (day)&70.000& $68.917^{+0.832}_{-0.985}$&$82.889^{+1.135}_{-1.087}$ &$80.745^{+1.352}_{-0.795}$    &${\cal U}_{\log} (0.000,200.000)$\\
10.     Decay duration ${\cal E}_1$ (day)&70.000& $72.996^{+2.941}_{-1.845}$&$75.709^{+0.792}_{-0.985}$ &$73.288^{+0.959}_{-0.693}$    &${\cal U}_{\log} (0.000,200.000)$\\
11.     Stable duration ${\cal L}_1$ (day) &30.000& $29.645^{+0.530}_{-0.798}$&$62.621^{+0.389}_{-0.462}$ &$62.315^{+0.337}_{-0.509}$ &   ${\cal U}_{\log}  (0.000,200.000)$\\
(\textit{2nd spot})&  &  & & &  \\
12.     Latitude $\Phi_2$  (deg) &30.00&$25.26^{+3.60}_{-0.46}$&$7.57^{+5.43}_{-2.69}$&$-0.29^{+0.62}_{-0.79}$     &${\cal U} (-90.00,90.00)$\\
13.     Initial longitude $\Lambda_2$ (deg) &-150.00  &$-151.20^{+1.49}_{-1.71}$&$-23.30^{+1.69}_{-1.51}$&$-22.40^{+2.19}_{-1.41}$ &${\cal U}  (-180.00,180.00)$\\
14.     Reference time $t_{{\scriptsize 2}}$ (day) &100.00&$99.89^{+0.29}_{-0.24}$& $151.95^{+0.43}_{-0.46}$&$152.90^{+0.39}_{-0.47}$   &${\cal U} (t_{{\scriptsize 1}},t_{{\scriptsize 3}})$\tablenotemark{d,e} \\
15.     Maximum radius $\alpha_{{\scriptsize \textrm{max,2}}}$ (deg)&5.00&$4.98^{+0.16}_{-0.21} $&$5.06^{+0.11}_{-0.20}$ &$5.67^{+0.21}_{-0.27}$    &${\cal U}(0.01,15.00)$\\
16.     Emergence duration ${\cal I}_2$ (day)&70.000 &$70.704^{+1.450}_{-1.497}$ &$64.258^{+1.167}_{-0.944}$&$70.520^{+0.873}_{-1.182}$   &${\cal U}_{\log} (0.000,200.000)$\\
17.     Decay duration ${\cal E}_2$ (day)&70.000 & $69.311^{+0.773}_{-0.854}$ &$49.009^{+5.633}_{-3.272}$&$41.850^{+5.284}_{-3.237}$  &${\cal U}_{\log} (0.000,200.000)$\\
18.     Stable duration ${\cal L}_2$ (day) &30.000&$30.472^{+0.420}_{-0.622}$&$34.450^{+0.928}_{-0.938}$ &$35.759^{+0.635}_{-1.175}$    &${\cal U}_{\log}  (0.000,200.000)$\\
(\textit{3rd spot}) &  &  & && \\
19.     Latitude $\Phi_3$  (deg) &15.00&$12.27^{+2.60}_{-0.59}$ &-&-     &${\cal U} (-90.00,90.00)$\\
20.     Initial longitude $\Lambda_3$ (deg) &-25.00 &$-23.50^{+1.39}_{-1.81}$ &-&-   &${\cal U}  (-180.00,180.00)$\\
21.     Reference time $t_{{\scriptsize 3}}$ (day) &150.00 &$150.51^{+0.39}_{-0.48}$&- &-  &${\cal U} (t_{{\scriptsize 2}},200.00)$\tablenotemark{d} \\
22.     Maximum radius $\alpha_{{\scriptsize \textrm{max,3}}}$ (deg)&5.00&$5.11^{+0.20}_{-0.19}$ &-&-   &${\cal U}(0.01,15.00)$\\
23.      Emergence duration ${\cal I}_3$ (day)&70.000 &$69.082^{+1.009}_{-1.028}$ &-&-   &${\cal U}_{\log} (0.00,200.000)$\\
24.     Decay duration ${\cal E}_3$ (day)&70.000 &$65.320^{+4.032}_{-4.379}$ &-&-  &${\cal U}_{\log} (0.00,200.000)$\\
25.     Stable duration ${\cal L}_3$ (day) &30.000 & $31.551^{+0.445}_{-1.434}$&-&-   &${\cal U}_{\log}  (0.00,200.000)$\\ \hline
Model evidence $\log{\cal Z}$ & &60257.571 &57617.691 &57507.414  & \\
\enddata
\tablenotetext{a}{
Each representation of the prior distributions defined in $a\le \theta \le b$ are as follows: bounded uniform distribution ${\cal U}(a,b)=1/(b-a)$; log uniform distribution ${\cal U_{\text{log}}}(a,b)= \log \theta/\log(b/a)$ known as Jeffey's prior; truncated normal distribution ${\cal TN} (\mu,\sigma^2,a,b)$, which equals ${\cal N}(\mu,\sigma^2)$ normalized by its cumulative distribution.}
\tablenotetext{b}{
The variance value is based on currently achieved precision of spectroscopy \citep{2019ApJ...876...58N}.
}
\tablenotetext{c}{
The variance value is adopted from a formula of the spot temperature based on the Doppler imaging technique in the case of solar effective temperature \citep{2005LRSP....2....8B,2017PASJ...69...41M}.}
\tablenotetext{d}{
We discern each of spot by the 
reference time $t_k$: if spots are not discerned, they are replaced one by one during the parallel tempering parameter transition. The number of maxima of the likelihood equals factorial of the number of spots, and the parallel tempering sampling becomes much inefficient.}
\tablenotetext{e}{
For the 2-spot model and 2-spot model with fixed $\sin i$, we set $t_3=200.000$ (upper limit of the interval of the light curve).
}
\end{deluxetable*}

\begin{deluxetable*}{lccccc}
\tablecaption{1-spot-like light curve case\label{tb:para2}}
\tabletypesize{\footnotesize}
\tablehead{
\colhead{Deduced parameters} & \colhead{Input} & \colhead{3-spot model} & \colhead{2-spot model} & \colhead{2-spot model}& \colhead{Prior distribution\tablenotemark{a}}  \\
\colhead{} &\colhead{value} &\colhead{} &\colhead{} & \colhead{with fixed $\sin i$}&\colhead{} 
}
\startdata
\textbf{(Stellar parameters)} &&&&&  \\
1.  Sine of inclination angle $\sin i$& 0.8660   & $0.8976^{+0.0054}_{-0.0647}$ & $0.9874^{+0.0013}_{-0.0024}$  & 0.8660 (\textit{fixed})   &${\cal TN}(0.8660,0.1500^2,0.0000,1.0000)$ \tablenotemark{b}\\
2.  Equatorial period $P_{{\scriptsize \textrm{eq}}}$ (day) & 25.0000 & $ 24.9756^{+0.0550}_{-0.0748}$&$24.4909^{+0.1400}_{-0.1002}$ &$25.0607^{+0.0251}_{-0.0526}$ &${\cal U}_{\log} (24.0000, 26.0000)$\\
3.   Degree of differential rotation $\kappa$ &0.1500 & $0.1340^{+0.0426}_{-0.0075}$ & $0.0467^{+0.0041}_{-0.0057}$  &$0.0331^{+0.0028}_{-0.0020}$ &${\cal U} (0.0000,0.2000)$\\
\textbf{(Spot parameters)} &  & &&&   \\
4.     Relative intensity $f_{\scriptsize \textrm{spot}}$ & 0.3000 & $0.3050^{+0.0487}_{-0.0336}$ & $0.2974^{+0.0767}_{-0.0152}$   &$0.3518^{+0.0486}_{-0.0609}$   &{${\cal TN} (0.3000,0.0500^2,0.1500,0.4500)$  \tablenotemark{c}}\\
(\textit{1st spot}) &  &  & && \\
5.     Latitude $\Phi_1$  (deg) &45.00& $48.28^{+2.20}_{-7.76}$ & $75.92^{+0.35}_{-0.72}$   &$55.38^{+0.22}_{-0.28}$   &${\cal U} (-90.00,90.00)$\\
6.     Initial longitude $\Lambda_1$  (deg)&55.00 & $54.90^{+0.69}_{-0.81}$  &$32.40^{+0.39}_{-0.51}$  &$32.60^{+0.49}_{-0.41}$   &${\cal U}  (-180.00,180.00)$\\
7.     Reference time $t_{{\scriptsize 1}}$ (day) & 50.00& $ 50.21^{+0.42}_{-0.52}$ & $83.86^{+0.13}_{-0.23}$ &$83.95^{+0.19}_{-0.18}$    &{${\cal U} (0.00,t_{{\scriptsize 2}})$\tablenotemark{d}} \\
8.     Maximum radius $\alpha_{{\scriptsize \textrm{max,1}}}$  (deg)& 5.00& $5.06^{+0.40}_{-0.17}$ & $11.34^{+0.30}_{-0.54}$  &$6.09^{+0.32}_{-0.20}$    &${\cal U} (0.01,15.00)$\\
9.     Emergence duration ${\cal I}_1$ (day)&70.000& $67.235^{+2.591}_{-3.003}$& $104.731^{+1.222}_{-1.042}$  &$101.301^{+1.131}_{-0.847}$    &${\cal U}_{\log} (0.000,200.000)$\\
10.     Decay duration ${\cal E}_1$ (day)&70.000& $69.930^{+4.191}_{-4.629}$&  $72.010^{+0.956}_{-1.975}$   &$67.904^{+1.130}_{-1.484}$    &${\cal U}_{\log} (0.000,200.000)$\\
11.     Stable duration ${\cal L}_1$ (day) &30.000& $30.624^{+0.754}_{-0.940}$&$61.412^{+0.400}_{-0.392}$    &$62.004^{+0.354}_{-0.426}$ &   ${\cal U}_{\log}  (0.000,200.000)$\\
(\textit{2nd spot})&  &  & & &  \\
12.     Latitude $\Phi_2$  (deg) &30.00&$34.08^{+1.61}_{-6.15}$& $45.43^{+2.24}_{-2.82}$  &$17.13^{+0.99}_{-1.00}$     &${\cal U} (-90.00,90.00)$\\
13.     Initial longitude $\Lambda_2$ (deg) &75.00  &$77.90^{+2.59}_{-2.21}$ &$-65.60^{+3.29}_{-2.51}$   &$-61.30^{+2.59}_{-3.51}$ &${\cal U}  (-180.00,180.00)$\\
14.     Reference time $t_{{\scriptsize 2}}$ (day) &100.00&$100.68^{+0.46}_{-0.56}$& $150.44^{+1.28}_{-0.93}$  &$150.95^{+0.93}_{-1.40}$   &${\cal U} (t_{{\scriptsize 1}},t_{{\scriptsize 3}})$\tablenotemark{d,e} \\
15.     Maximum radius $\alpha_{{\scriptsize \textrm{max,2}}}$ (deg)&5.00&$5.05^{+0.17}_{-0.14}$ &$5.57^{+0.30}_{-0.14}$    &$4.99^{+0.21}_{-0.22}$    &${\cal U}(0.01,15.00)$\\
16.     Emergence duration ${\cal I}_2$ (day)&70.000 &$66.746^{+3.542}_{-2.173}$&$39.158^{+0.775}_{-0.940}$  &$38.637^{+0.867}_{-0.852}$   &${\cal U}_{\log} (0.000,200.000)$\\
17.     Decay duration ${\cal E}_2$ (day)&70.000 & $69.651^{+1.272}_{-1.767}$&$100.786^{+6.404}_{-17.922}$  &$97.018^{+8.629}_{-16.458}$  &${\cal U}_{\log} (0.000,200.000)$\\
18.     Stable duration ${\cal L}_2$ (day) &30.000&$30.963^{+0.755}_{-0.764}$&$25.394^{+1.430}_{-3.138}$  &$24.693^{+1.856}_{-3.006}$    &${\cal U}_{\log}  (0.000,200.000)$\\
(\textit{3rd spot}) &  &  & && \\
19.     Latitude $\Phi_3$  (deg) &15.00&$15.98^{+1.64}_{-3.22}$ &-&-     &${\cal U} (-90.00,90.00)$\\
20.     Initial longitude $\Lambda_3$ (deg) &-50.00 &$-54.00^{+4.49}_{-2.21}$ &-&-   &${\cal U}  (-180.00,180.00)$\\
21.     Reference time $t_{{\scriptsize 3}}$ (day) &150.00 &$149.90^{+0.86}_{-1.37}$&- &-  &${\cal U} (t_{{\scriptsize 2}},200.00)$\tablenotemark{d} \\
22.     Maximum radius $\alpha_{{\scriptsize \textrm{max,3}}}$ (deg)&5.00&$5.04^{+0.15}_{-0.18} $ &-&-   &${\cal U}(0.01,15.00)$\\
23.      Emergence duration ${\cal I}_3$ (day)&70.000 &$70.000^{+3.545}_{-5.136}$ &-&-   &${\cal U}_{\log} (0.00,200.000)$\\
24.     Decay duration ${\cal E}_3$ (day)&70.000 &$71.902^{+15.202}_{-6.103} $ &-&-  &${\cal U}_{\log} (0.00,200.000)$\\
25.     Stable duration ${\cal L}_3$ (day) &30.000 & $29.131^{+1.662}_{-3.178}$&-&-   &${\cal U}_{\log}  (0.00,200.000)$\\ \hline
Model evidence $\log{\cal Z}$ & &60307.348 & 59406.380 &59355.393  & \\
\enddata
\tablenotetext{a}{
Each representation of the prior distributions defined in $a\le \theta \le b$ are as follows: bounded uniform distribution ${\cal U}(a,b)=1/(b-a)$; log uniform distribution ${\cal U_{\text{log}}}(a,b)= \log \theta/\log(b/a)$ known as Jeffey's prior; truncated normal distribution ${\cal TN} (\mu,\sigma^2,a,b)$, which equals ${\cal N}(\mu,\sigma^2)$ normalized by its cumulative distribution.}
\tablenotetext{b}{
The variance value is based on currently achieved precision of spectroscopy \citep{2019ApJ...876...58N}.
}
\tablenotetext{c}{
The variance value is adopted from a formula of the spot temperature based on the Doppler imaging technique in the case of solar effective temperature \citep{2005LRSP....2....8B,2017PASJ...69...41M}.}
\tablenotetext{d}{
We discern each of spot by the  
reference time $t_k$: if spots are not discerned, they are replaced one by one during the parallel tempering parameter transition. The number of maxima of the likelihood equals factorial of the number of spots, and the parallel tempering sampling becomes much inefficient.}
\tablenotetext{e}{
For the 2-spot model and 2-spot model with fixed $\sin i$, we set $t_3=200.000$ (upper limit of the interval of the light curve).
}
\end{deluxetable*}

\end{document}